\documentclass[aps,prb,twocolumn,showpacs,superscriptaddress,floatfix]{revtex4-2}  
\usepackage{ucs}
\usepackage{natbib}
\usepackage{graphicx}  % needed for figures
\usepackage{bm}        % for math
\usepackage{amssymb}   % for math
\usepackage{amsmath}
\usepackage{color}
\usepackage{CJK}
\usepackage{times}
\usepackage{verbatim}
\usepackage{mathrsfs}
\usepackage{hyperref}
\usepackage[normalem]{ulem}
\usepackage{physics}
\hypersetup{colorlinks = True, urlcolor=blue, linkcolor=blue, citecolor=blue}
\newcommand{\bk}{\mathbf{k}}
\newcommand{\bp}{\mathbf{p}}
\newcommand{\bq}{\mathbf{q}}
\newcommand{\br}{\mathbf{r}}

\newcommand{\sig}[2]{\sigma^{#1}_{#2}}
\newcommand{\bg}{\begin{pmatrix}}
\newcommand{\ed}{\end{pmatrix}}
\newcommand{\dg}{\dagger}
\newcommand{\sg}{\sigma}
\newcommand{\al}{\alpha}
\newcommand{\bt}{\beta}
\newcommand{\kp}{\kappa}
\newcommand{\ld}{\lambda}
\newcommand{\Ld}{\Lambda}
\newcommand{\gm}{\gamma}
\newcommand{\ep}{\epsilon}
\newcommand{\dt}{\delta}

\newcommand{\pair}[2]{\langle #1 \rangle_{#2}}

\usepackage{cancel}
\usepackage{comment}

\usepackage{newtxmath}
\usepackage{comment}
\hypersetup{colorlinks = True, urlcolor=blue, linkcolor=blue, citecolor=blue}

\newcommand{\Dt}{\Delta}
\newcommand{\zt}{\zeta}

\newcommand{\calH}{\mathcal{H}}
\newcommand{\calV}{\mathcal{V}}
\newcommand{\calI}{\mathcal{I}}
\newcommand{\calX}{\mathcal{X}}
\newcommand{\calY}{\mathcal{Y}}
\newcommand{\calZ}{\mathcal{Z}}

\newcommand{\bdt}{\bm{\delta}}
\newcommand{\fc}[1]{d^{\dg}_{#1}}
\newcommand{\fa}[1]{d^{\phantom{\dg}}_{#1}}

\newcommand{\om}{\omega}
\newcommand{\Om}{\Omega}

\newcommand{\bA}{\mathbf{A}}

\newcommand{\ba}{\mathbf{a}}

\newcommand{\bK}{\mathbf{K}}
\newcommand{\bG}{\mathbf{G}}

\newcommand{\bR}{\mathbf{R}}

\newcommand{\bro}{\bm{\rho}}

\newcommand{\et}{\eta}

\newcommand{\calP}{\mathcal{P}}
\newcommand{\calQ}{\mathcal{Q}}

\begin{document}

\title{Flux-induced Aharonov-Bohm Oscillation in the Tunneling Spectroscopy of Kitaev Spin Liquids}
%\input author_list.tex       
% D0 authors (remove the first 3 lines
% of this file prior to submission, they
% contain a time stamp for the authorlist)
% (includes institutions and visitors)
\author{Wen-Han Kao}
%\email{}
%\affiliation{}
\author{Elio J. König}
\affiliation{Department of Physics, University of Wisconsin--Madison, Madison, WI 53706, USA}

%\date{\today}
\begin{abstract}
Identifying emergent flux excitations and signatures of their mutual statistics with itinerant Majorana fermions remains a central challenge in Kitaev spin liquids. We show that an isolated $\pi$ flux imprints an intrinsic Aharonov–Bohm-like interference pattern on the charge-neutral Majorana continuum. Spatially resolved scanning tunneling microscopy with inelastic electron tunneling spectroscopy (STM-IETS) probes this oscillation in the local dynamical spin response as a function of both bias voltage and tip–flux separation, reflecting the $\pi$ phase acquired by Majorana trajectories encircling the flux. We show that large-scale exact diagonalization of the Kitaev honeycomb model and a complementary low-energy continuum theory yield consistent radial oscillations. These results provide a finite-energy, spatially resolved tunneling signature of a preexisting flux and a braiding-related diagnostic of fermion-flux mutual statistics without requiring manipulation of individual anyons in the Kitaev spin liquid.
\end{abstract}
\maketitle

\section{Introduction}

Quantum spin liquids (QSLs) are among the most profoundly quantum states of matter. The term refers to quantum-disordered phases of interacting
spin systems that evade conventional magnetic long-range order even in the
zero-temperature limit \cite{Anderson1973, Balents2010, Savary2017, Zhou2017, Broholm2020}. However, the absence of magnetic order alone is not sufficient to establish
a QSL because similar phenomenology is also exhibited in trivial paramagnets, spin glasses, and strong-disorder magnetic states. Instead, QSLs are distinguished
by long-range many-body entanglement, emergent gauge structure, and
deconfined fractionalized excitations. In gapped QSLs, these properties are
encoded in intrinsic topological order and anyonic quasiparticles. In contrast, gapless QSLs are more generally characterized by gapless fractionalized
degrees of freedom coupled to emergent gauge fields. Because these defining
features are intrinsically nonlocal and many observable signatures are
indirect or non-unique, identifying a QSL generally requires consistent
evidence from several complementary probes \cite{KnolleMoessner2019}.

A paradigmatic
analytically tractable Hamiltonian for a QSL
is the Kitaev honeycomb model, in which quantum spins fractionalize into itinerant Majorana fermions coupled to an emergent $\mathbb{Z}_2$ gauge field and the fermionic Hamiltonian is exactly solvable \cite{Kitaev2006, Knolle2017ARCMP, Motome2019, Nasu2024, Kee2025RMP}. Depending on the exchange couplings, this model supports gapped Abelian $\mathbb{Z}_2$ spin-liquid phases as well as a gapless spin-liquid phase. An exactly solvable time-reversal-breaking three-spin perturbation, generated perturbatively by a weak magnetic field, gaps the Majorana Dirac cones and induces a chiral non-Abelian phase with Ising topological order and a chiral Majorana edge mode.

\begin{figure}
\includegraphics[width=1.0\columnwidth]{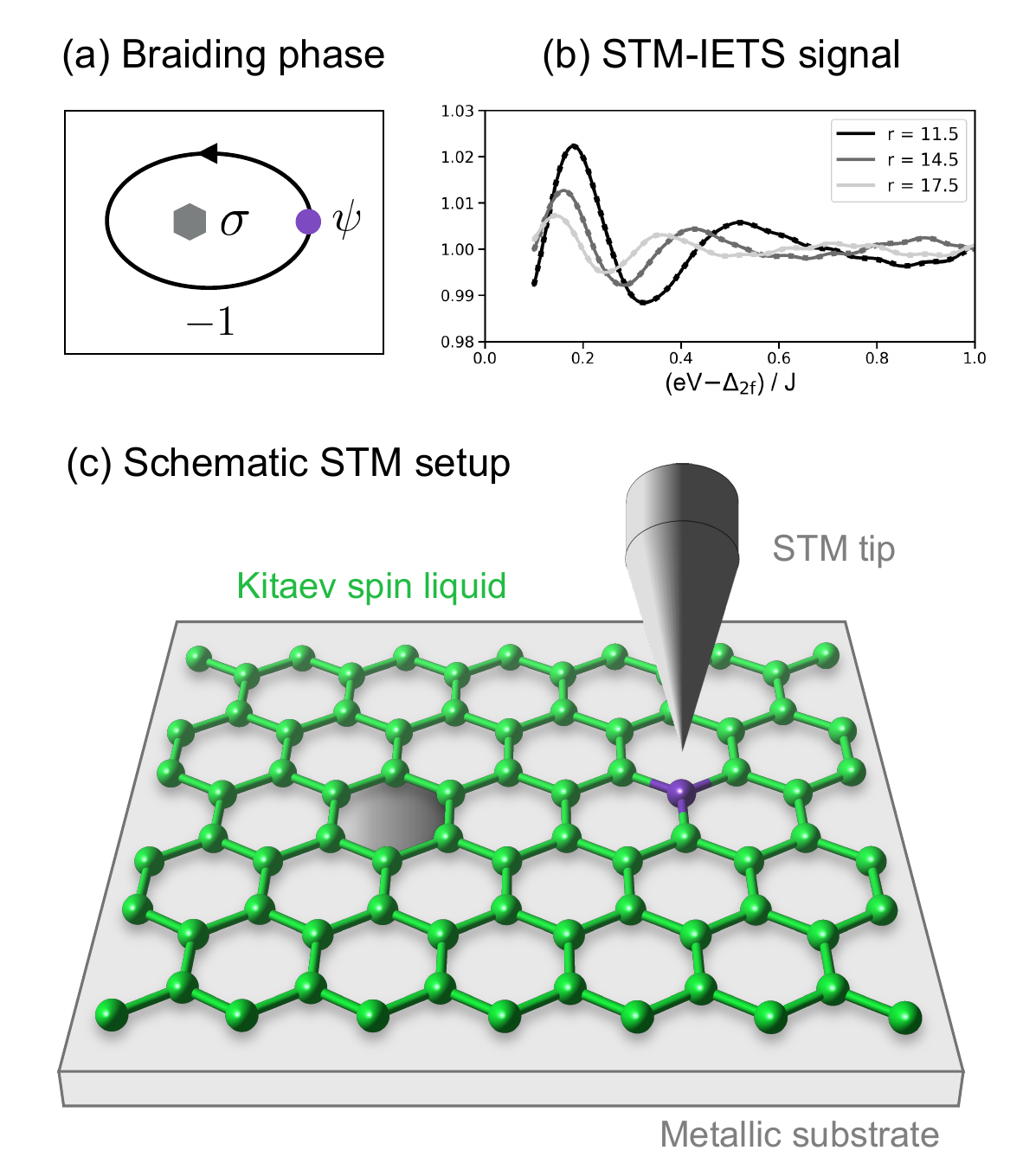}
     \caption{
     Proposed scanning tunneling microscopy (STM) setup for probing the Ising anyon braiding effect from an isolated $\pi$-flux excitation. {(a) The full braiding phase from a fermion encircling an isolated $\pi$-flux excitation. (b) The proposed tunneling signal as a function of the bias voltage $eV$ (subtracted by the two-flux gap $\Dt_{2f}$), in units of the Kitaev coupling $J$. Different curves represent the different measuring positions with a horizontal distance $r$ in units of the lattice constant from the $\pi$-flux plaquette. (c) The schematic STM setup, where the tip is on top of one lattice site (highlighted in purple) in the Kitaev spin-liquid layer. The isolated $\pi$-flux excitation is depicted as the shaded hexagon.}}
     \label{Fig:schematic_setup}
\end{figure}

Surprisingly, the bond-directional exchange structure of the Kitaev model is not merely
a mathematical construction. In the spin-orbit-assisted Mott insulators with
edge-sharing ligand octahedra, the interplay of strong spin-orbit coupling,
electron correlations, and orbital-dependent superexchange can generate
Kitaev interactions between spin-orbit-entangled local moments \cite{Jackeli2009, Rau2014, Rau2016, Winter2017b, Takagi2019, Trebst2022, Ojeda-Aristizabal2026}. Note that real candidate materials generally contain additional Heisenberg, off-diagonal, and longer-range exchanges, and are
therefore more accurately described as proximate Kitaev magnets than as
realizations of the pure Kitaev Hamiltonian. Establishing an unambiguous material realization of a Kitaev spin liquid consequently remains an active experimental challenge.

In the chiral non-Abelian phase of the Kitaev spin liquid, the $\mathbb{Z}_2$ flux excitations realize Ising-type non-Abelian anyons, making this phase a promising platform for topological quantum computation \cite{Lahtinen2009,Lahtinen2011}. A central challenge, however, is that both the spin-liquid phase itself and the associated Ising anyons are difficult to identify unambiguously in experiment, because conventional probes couple only to gauge-invariant spin operators rather than directly to the fractionalized quasiparticles. The defining statistical properties of non-Abelian anyons are encoded in their fusion channels and braiding operations \cite{Kitaev2003, Nayak2008}. Although fusion-based signatures may provide particularly direct evidence, most existing proposals require the ability to create, detect, and manipulate individual flux excitations, which remains experimentally demanding \cite{Lahtinen2009, Alicea2020, Klocke2024, Principi2026}. A complementary and potentially more accessible route is to search for a braiding-related interference effect in the presence of a fixed, isolated flux. In this setting, an itinerant Majorana fermion encircling the $\mathbb{Z}_2$ flux acquires a mutual braiding phase of -1; see Fig.~\ref{Fig:schematic_setup}(a). The on-site Majorana Green’s function offers a natural diagnostic of this mutual braiding effect. 

The accumulation of a braiding phase becomes particularly evident in a path-integral description, where the low-energy Green's function can be obtained by summing over closed semiclassical fermion trajectories; see App.~\ref{app:Feynman}. 
Trajectories that wind around the flux an odd number of times acquire an additional minus sign in the amplitude. This produces an Aharonov-Bohm-like quantum oscillation~\cite{ABeffect1959} in the Majorana local density of states, analogous to electronic interference around an external magnetic flux solenoid, with a crucial distinction that the flux here is an intrinsic fractionalized excitation of the spin liquid.

In this work, we investigate how such an intrinsic Aharonov-Bohm oscillation from an isolated $\pi$-flux excitation can appear in experimentally relevant local spin spectroscopy. In the STM-IETS (Scanning Tunneling Microscopy with Inelastic Electron Tunneling Spectroscopy) setup as shown in Fig.~\ref{Fig:schematic_setup}(c), the tunneling conductance can probe the local dynamical spin correlation function of the Kitaev layer, which contains information about the underlying fermionic quasiparticle spectrum \cite{Konig2020, Knolle2020, Udagawa2021, Bauer2023, takahashi2023nonlocal, Kao2024PRL, Kao2024PRB, Bauer2024, Zhang2025_PRB, Zhang2025_npj} (see Sec.~\ref{Sec:KSL_IETS}, below, for a summary). 
The braiding-induced oscillations are then experimentally accessible by normalizing the inelastic tunneling signal $\mathrm{d}^2I(r)/\mathrm{d}V^2$ at distance $r$ from a $\pi$-flux excitation with the signal in the pristine environment $\mathrm{d}^2I(r\to \infty)/\mathrm{d}V^2$; see Fig.~\ref{Fig:schematic_setup}(b) \footnote{This proposed signal is calculated using real-space lattice exact diagonalization for an $L=122$ system with shifted boundary conditions, similar to Fig.~\ref{Fig:STM_interval3}(b). Here, however, we adopt a different normalization scheme and present the data as the ratio of $\mathrm{d}^2I(r)/\mathrm{d}V^2$ to $\mathrm{d}^2I(r\to\infty)/\mathrm{d}V^2$, which is more readily accessible experimentally. For the other figures, we normalize the signal with respect to the linear dispersion, as this facilitates comparison with continuum-model calculations.}.

Unlike a bare Majorana Green’s function, however, a local spin flip in the spin-spin correlation function also creates a nearby pair of $\mathbb{Z}_2$ fluxes at the tunneling position. This flux-pair insertion acts as a local quench for the Majorana fermions and can generate resonance features in the response, but does not destroy the oscillatory feature from the distant isolated $\pi$-flux excitation. We reveal and analyze this phenomenon by using two complementary approaches. First, we compute the STM-IETS response of the lattice Kitaev model by exact diagonalization in the presence of an isolated flux. Second, we derive a low-energy continuum description in terms of massive Dirac Majorana fermions coupled to a flux-solenoid-like vector potential, and treat the additional flux-pair effect as an effective local tip potential in the  T-matrix formalism. The two approaches yield consistent Aharonov-Bohm-like signatures, indicating that the oscillation observed in the lattice calculation is a robust physical consequence of Majorana braiding around an intrinsic $\mathbb{Z}_2$ flux, rather than an artifact of finite-size degeneracies or the discreteness of the numerical spectrum.

The remainder of this paper is organized as follows. In Sec.~\ref{Sec:KSL_IETS}, we introduce the Kitaev honeycomb model, and review the local dynamical spin correlations and their connection to STM-IETS signals. In Sec.~\ref{Sec:AB_oscillation}, we describe the lattice-calculation setup and present exact-diagonalization results demonstrating the Aharonov-Bohm-like oscillation induced by an isolated flux. In Sec.~\ref{Sec:continuum_model}, we derive the wavefunctions of the effective Dirac Hamiltonian in the presence of a flux-induced vector potential and analyze the resulting Aharonov-Bohm oscillations, both without and with the effective local potential induced by the flux-pair insertion. In the appendices, we provide technical details on tunneling spectroscopy in App.~\ref{app:tunneling}, the derivation of the low-energy continuum model and its solution in App.~\ref{app:low_E_theory}, the derivation of the retarded Green's function in the zero-flux and $\pi$-flux sectors in App.~\ref{app:derive_GF}, their low-energy limit in App.~\ref{app:LowEnergyGF}, and the semiclassical calculation from the Feynman path-integral approach in App.~\ref{app:Feynman}.

\section{Local dynamical response of the Kitaev spin liquid}\label{Sec:KSL_IETS}

In this section, we review the most salient aspects of Kitaev quantum spin liquids, their dynamical spin response, and how the latter can be probed in tunneling spectroscopy.

\subsection{Hamiltonian and real-space diagonalization}

The lattice Hamiltonian considered in this work is the Kitaev honeycomb model~\cite{Kitaev2006} with a three-spin perturbation term that breaks the time-reversal symmetry and opens a Haldane-like mass gap on the Dirac cones~\cite{Haldane1988}. The Kitaev interactions are considered isotropic and $J = 1$ throughout the paper:
\begin{align}\label{Eq:Kitaev_Hamiltonian}
\mathcal{H} &= -J\sum_{\pair{jk}{\al}}\sg^{\al}_{j}\sg^{\al}_{k}-\kappa \sum_{\langle jkl\rangle_{\al\bt}}\sg^{\al}_{j}\sg^{\gm}_{k}\sg^{\bt}_{l}.
\end{align}
In Kitaev's four-Majorana parton construction, the spin-1/2 operators are decomposed into Majorana operators $\sig{\al}{j} = ib^{\al}_{j}c^{\phantom{\al}}_{j}$ with $\{b^{\al}_i, b^{\bt}_j\} = 2\dt^{\al\bt}\dt_{ij}$ and $\{c_i,c_j\} = 2\dt_{ij}$. The $b$-Majorana operators on the same bond form a static $\mathbb{Z}_2$ gauge field $u_{\pair{jk}{\al}} = ib^{\al}_{j}b^{\al}_{k}$ and the Hamiltonian becomes bilinear in the Majorana representation
\begin{align}
&\mathcal{H} = iJ\sum_{\pair{jk}{\al}}u_{\pair{jk}{\al}}c_j c_k+i\kappa\sum_{\langle jkl\rangle_{\al\bt}}u_{\pair{jk}{\al}}u_{\pair{kl}{\bt}}c_j c_l. \label{eq:HMajo}
\end{align}
In this representation, the model becomes a free-fermion Hamiltonian with first-neighbor hopping $J$ and the chiral second-neighbor hopping $\kp$. The local conserved flux operator defined on each hexagon of the lattice can be calculated from the $\mathbb{Z}_2$ gauge fields:
\begin{align}
W_p = \prod_{\langle jk \rangle_{\alpha} \in p} u_{\langle jk \rangle_{\alpha}} = \pm 1,
\end{align}
where $W_p = +1$ denotes no flux or \textit{zero flux} on a plaquette, and $W_p = -1$ denotes a \textit{$\pi$-flux} excitation.
 
Due to the bipartite nature of the honeycomb lattice, the real-space Hamiltonian can be written in the sublattice space and then transformed into a Bogoliubov-de Gennes form of complex fermions via
\begin{align}
\bg f \\ f^{\dg} \ed = \frac{1}{2}\bg 1 & i \\ 1 & -i \ed \bg c_A \\ c_B \ed,
\end{align}
and the resulting Hamiltonian is
\begin{align}
\begin{split}
\mathcal{H} &= \frac{1}{2} \bg f^{\dg} & f \ed \bg \tilde{h} & \tilde{\Delta} \\ \tilde{\Delta}^{\dg} & -\tilde{h}^{T} \ed \bg f \\ f^{\dg} \ed, \label{eq:HSC}
\end{split}
\end{align}
where $\tilde{h}$ and $\tilde{\Dt}$ are normal and pairing blocks, respectively.

Following Ref.~\cite{Knolle2014,Knolle2015}, the Hamiltonian can be diagonalized by using the Bogoliubov transformation
\begin{align}
\begin{split}
&\mathcal{H} = \sum_{n}\epsilon_n \left( a^{\dg}_n a_n - \frac{1}{2} \right)
\end{split}
\end{align}
with the excitation energies $\epsilon_n > 0$. Therefore, the ground-state energy for a given static flux sector is $E_{0}(\{u\}) = -\frac{1}{2}\sum_n \epsilon_n$. In the limit of small $\kappa/J$, Lieb's theorem ~\cite{Lieb1994} and adiabaticity ensure the ground state resides in the flux-free sector {with $W_p = +1$ for all $p$}, and $\epsilon_n$ can readily be calculated by using the Fourier transform {(see App.~\ref{app:EffectiveU1Theory})}.

\subsection{Dynamical spin correlation function}\label{Sec:dynamical_spin_correlation}

Following Ref.~\cite{Baskaran2007,Tikhonov2011,Knolle2014,Knolle2015,Kao2024PRL,Kao2024PRB}, we derive the dynamical spin correlation function for the Kitaev honeycomb model, which can be measured in the STM-IETS signal. The general expression for the dynamical spin correlation in the ground state is
\begin{align}
S^{\al\bt}_{jk}(t) &= \bra{0}\sg^{\al}_j (t)\sg^{\bt}_k (0)\ket{0}.
\end{align}
However, the orthogonality of different flux sectors restricts the correlations to be ultra-short-ranged, that is, only the same site ($S^{\al\al}_{jj}$) or nearest-neighbor sites with the same spin component ($S^{\al\al}_{\langle jk\rangle_{\al}}$) have a nonzero correlation. The integrability of the Kitaev model implies that all states can be decomposed into the flux sector and the matter-fermion sector, in particular we denote the ground state as $\ket{0} = \ket{F}\otimes \ket{M}$, and the dynamical correlation can be rewritten by using the bond-fermion approach~\cite{Knolle2015,Kao2024PRB}. In this work, we focus on the on-site dynamical correlation with the form
\begin{align}
S^{\al\al}_{jj}(t) = e^{iE_0 t}\bra{M}c_j e^{-i\mathcal{H}^{\prime}t}c_j\ket{M},
\end{align}
where $E_0$ is the ground-state energy in the zero-flux sector, and $\calH'$ is the Hamiltonian with a flip of the link variable $u_{\langle jk\rangle_\alpha} \to -u_{\langle jk\rangle_\alpha}$, leading to a {flux-pair insertion} at the measuring site. The corresponding matter-fermion ground state in this two-flux sector is denoted as $\ket{M'}$, and the corresponding ground-state energy is $E_0'$. We employ the Lehmann spectral representation for $\calH'$ and {use the adiabatic approximation $\ket{M} \to \ket{M'}$}:
\begin{align}
\begin{split}
S^{\al\al}_{jj}(t) &\simeq \sum_{\ld}e^{i(E_0-E_{\ld}') t} \bra{M'}c_j\ket{\ld}\bra{\ld}c_j\ket{M'}\\
&= \sum_{\ld}e^{i(E_0-E_{\ld}') t} \bra{M'}c_j(a'_{\ld})^{\dg}\ket{M'}\bra{M'}a'_{\ld}c_j\ket{M'} \label{eq:Stspace}
\end{split}
\end{align}
where we use $\ket{\ld} = (a^{\prime}_{\ld})^{\dg}\ket{M^{\prime}}$ and $\mathcal{H}^{\prime}\ket{\ld} = E'_{\ld}\ket{\ld}$.  
This eigenenergy contains the ground-state energy and the one-particle excitation energy, $E'_{\ld} = E'_0+\ep'_{\ld}$. {The adiabatic approximation builds on the absence of an X-ray edge catastrophe for (pseudo-)gapped matter density of states and has been shown to give quantitatively good results compared with the exact dynamical correlations \cite{Knolle2014, Knolle2015, Konig2020}. }
In the frequency domain, it becomes
\begin{align}\label{Eq:onsite_spin_correlation}
\begin{split}
&S^{\al\al}_{jj}(\omega) = \sum_{\ld}|\bra{M'}a'_{\ld}c_j\ket{M'}|^2 \dt(\om-\ep'_{\ld}-\Dt_{2f}),
\end{split}
\end{align}
where the resonance frequency involves the two-flux excitation gap $\Delta_{2f} \equiv E^{\prime}_0 - E_0$. It is important to note that this expression can be interpreted as the local density of states (LDOS) of the Majorana fermions in the presence of the flux-pair insertion around the measuring site and an additional spectral gap $\Dt_{2f}$. When the measuring site is far away from other defects of the system, $\Dt_{2f} \approx 0.26J \times [1 + \mathcal O(\kappa/J)]$ is a position-independent constant \cite{Kitaev2006, Feng2020,Panigrahi2023}.

{In Sec.~\ref{Sec:AB_oscillation}, we present the numerical results of $S^{\al\al}_{jj}(\om)$ based on the lattice exact diagonalization of the Kitaev honeycomb model.} In Sec.~\ref{Sec:continuum_model}, we will consider the low-energy theory of Dirac fermions in which a fixed isolated $\pi$ flux at the origin gives rise to Aharonov-Bohm oscillation in the on-site LDOS. In that description, the local flux-pair insertion required in the dynamical spin correlation function of the Kitaev model can be considered as an effective local potential, and the full on-site LDOS calculated in the T-matrix approach gives the same physical quantity as Eq.~(\ref{Eq:onsite_spin_correlation}):
\begin{align}
S^{\al\al}_{jj}(\om)  \longleftrightarrow \rho^{\rm full}(\br_j, \om-\Dt_{2f}). \label{eq:DOS_DynSuscept}
\end{align}

\subsection{Tunneling spectroscopy}
\label{sec:tunneling}

We consider the inelastic electron tunneling spectroscopy with a sharp STM tip to probe the dynamical spin correlation of the Kitaev spin liquid, see Fig.~\ref{Fig:schematic_setup}, {where the inelastic contributions stem from cotunneling events ~\cite{Anderson1966, Appelbaum1966, Appelbaum1967,Rossier2009, Fransson2010}.} Deferring a detailed derivation to App.~\ref{app:tunneling}, we write the zero-temperature tunneling current $I = I_{\rm el} + I_{\rm inel}$ as
composed of elastic
\begin{subequations}
\begin{equation}\label{eq:IelMaintext}
   I_{\rm el} = G V
\end{equation}
and inelastic 
\begin{equation} \label{eq:IinelMaintext}
    I_{\rm inel} = 2\pi e t_1^2\sum_{ss'}D^{\rm tip}_{s'}D^{\rm sub}_{s}\sum_{\al} \vert\tau^{\al}_{ss'} \vert^2 \int_0^{eV}\mathrm{d}\om (eV-\om)S^{\al\al}_{jj}(\om)
\end{equation}
contributions ~\cite{Konig2020, Knolle2020, Udagawa2021, Bauer2023, takahashi2023nonlocal, Kao2024PRL, Kao2024PRB, Zhang2025_PRB, Zhang2025_npj}. Here, $G$ is the Ohmic conductance, $V$ the bias voltage, $t_1$ the cotunneling amplitude associated with spin-flipping processes, $D_s^{\rm tip},(D_s^{\rm sub})$ the density of states at the Fermi level of electrons with spin-projection $s$
in the tip (substrate) and $\tau^\alpha$ ($\alpha = x,y,z$) Pauli matrices in spin space.
\end{subequations}

We conclude that the STM-IETS signal, or the second derivative of the tunneling current, is proportional to the on-site spin-spin correlation function of the Kitaev spin liquid:
\begin{align}
\frac{\mathrm{d}^2 I}{\mathrm{d} V^2} \sim \sum_{\al}S^{\al\al}_{jj}(eV). \label{eq:IinelMaintextSimple}
\end{align}

Before closing this section, we comment on the distinction between the STM-IETS setup and other tunneling spectroscopic methods proposed for Kitaev spin liquids. For example, the planar tunneling junction can efficiently detect spatially integrated spectral weight and, in the presence of vacancies, can still reveal the low-energy characteristics of the vacancy-induced Majorana modes \cite{Carrega2020, Konig2020, Li2026_planar}. However, the Aharonov-Bohm oscillations studied in this work are encoded in the spatial evolution of the phase and amplitude of the local response around the isolated flux.  Spatial integration required by the planar-junction setup generally averages over different portions of the oscillatory pattern and suppresses this spatial dependence. The scanning capability of STM-IETS is essential for
imaging the radial flux-induced interference pattern. 

{Our proposal is also physically distinct from the recent STM experiment on the monolayer $\al$--RuCl$_3$, where the tunneling signature is observed at a bias exceeding the charge gap ~\cite{Kohsaka2024}. A related quasiparticle interference (QPI) mechanism has been discussed in a recent theoretical paper~\cite{Jahin2025}, where -- unlike in our setup -- the authors consider a direct electron addition into the Mott-insulating Kitaev layer.} The injected electron fractionalizes into a real chargon and a spinon, and the measured electron LDOS is consequently a convolution of the charge and spin sectors. In suitable bandwidth regimes, that convolution can be used to extract the spinon LDOS and dispersion from impurity-induced QPI. {The idea of STM electron tunneling into the Hubbard band has been discussed in different strongly correlated systems ~\cite{Mross2011,He2023, Wang2026}. } In contrast, the STM-IETS mechanism studied here transfers an electron between the STM tip and a metallic substrate through the magnetic layer, while leaving no real charge excitation in the Kitaev spin-liquid layer. Experimentally, this cotunneling setup has been applied to several 2D van der Waals materials to investigate their magnetic properties ~\cite{Klein2018, Ghazaryan2018, Kim2019, Yang2023, Ganguli2023, Zhang2025}. 

\section{Flux-induced Aharonov-Bohm oscillation}\label{Sec:AB_oscillation}

\begin{figure*}
\includegraphics[width=1.0\textwidth]{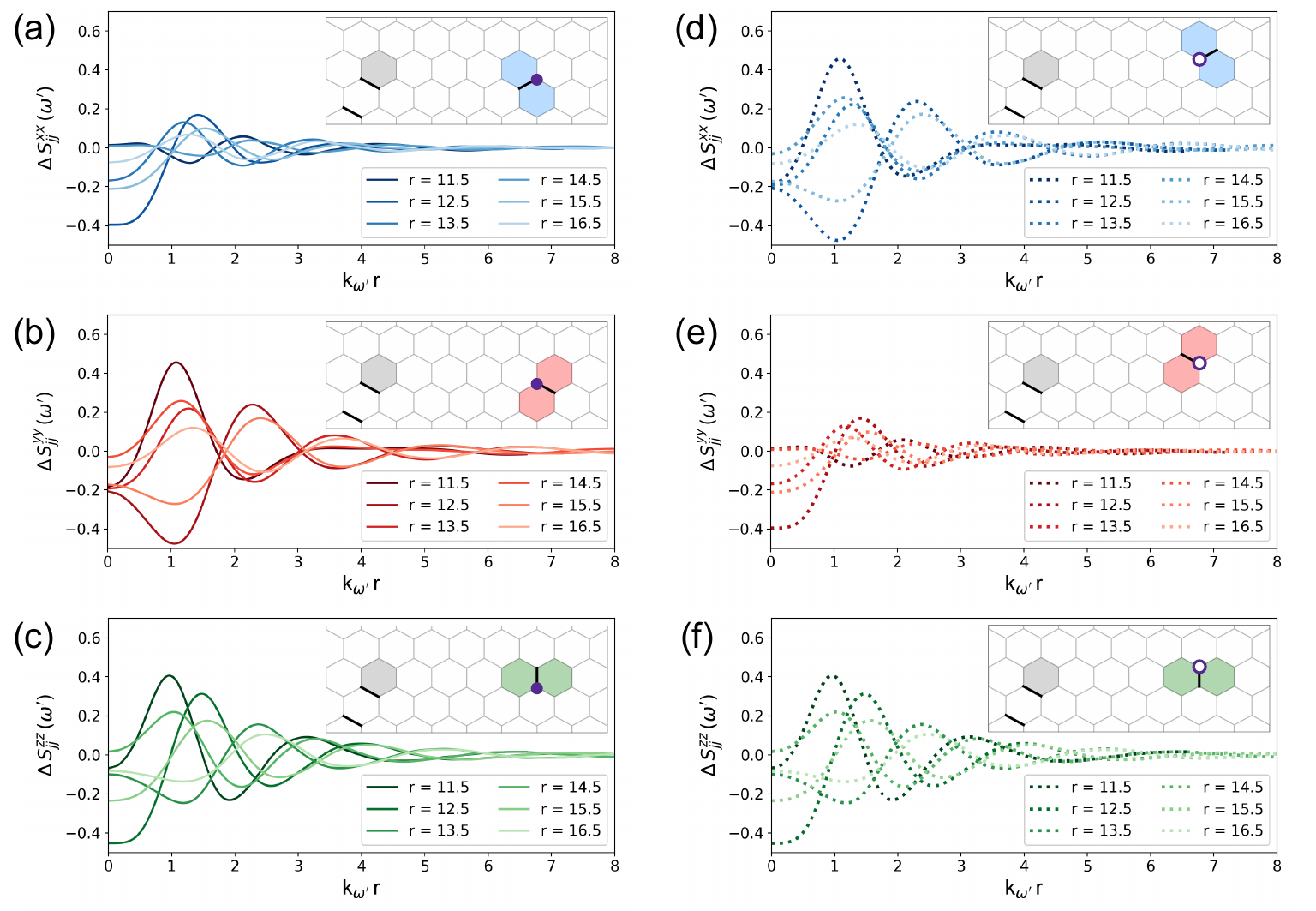}
     \caption{\label{Fig:STM_interval1} Different components of the flux-contrasted dynamical spin correlation function for (a)-(c) the A sublattice sites and (d)-(f) the B sublattice sites. {On the horizontal axis, we define $k_{\om'} = \sqrt{(\om')^2-\Dt^2}/v_s$ and $\om' = \om - \Dt_{2f}$}, where $\Dt_{2f}$ is the constant two-flux gap, $\Dt$ is the fermionic gap $\Dt = 6\sqrt{3}\kp$ of the Kitaev model, and $v_s = \sqrt{3}$ is the Fermi velocity. We keep $r$ fixed for each curve.    
     The inset of each panel depicts the measuring site (as a solid circle for A sublattice or a hollow circle for B sublattice) and the corresponding flux sector. The gray hexagon represents the isolated $\pi$ flux at the center of the lattice, which is the source of the Aharonov-Bohm oscillation. The colored hexagons are the fluxes created by the spin operator in the correlation function. The black thick lines denote the flipped gauge fields $u_{\langle jk\rangle_{\al}} = -1$ to create the flux sector. Note that the measuring point on the inset has a distance of $r = 4.5$ unit cells from the center of the gray hexagon, but the data shown in the main panel are for larger distances where the flux-flux interaction is suppressed.}
\end{figure*}

\begin{figure*}
\includegraphics[width=1.0\textwidth]{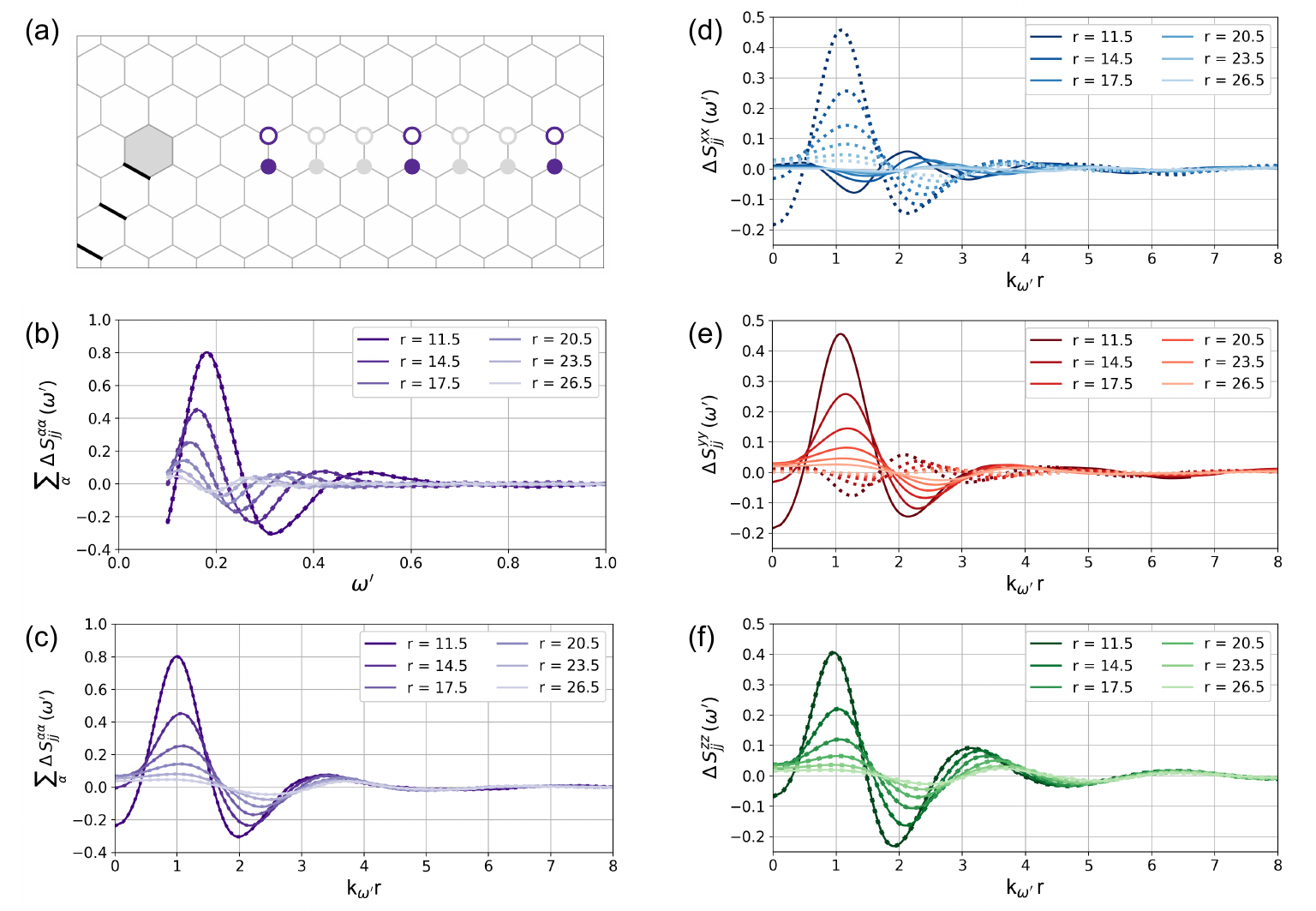}
     \caption{\label{Fig:STM_interval3} STM-IETS signals at measurement sites separated by three unit cells. (a) The setup for the measuring sites. Instead of measuring every site along the radial direction  ($\Dt r = 1$), we select sites separated by $\Dt r = 3$ to remove the relative intervalley phase in the oscillation. (b)-(c) Aharonov-Bohm oscillations in the STM-IETS signal as a function of frequency or $k_{\om'} r$. Note that we shift the frequency with the two-flux gap $\Dt_{2f}$ to show the alignment of the peaks. (d)-(f) Three components of the STM-IETS signals under the same $\Dt r = 3$ selection.}
\end{figure*}

This section contains a numerical evaluation of the local on-site dynamical spin-correlation function entering the inelastic tunneling current Eq.~\eqref{eq:IinelMaintextSimple}.

\subsection{Numerical implementation}

In the numerical calculation, we diagonalize the above real-space fermionic tight-binding lattice model for $L$ unit cells in each dimension, such that the total number of lattice sites is $N = 2L^2$, including both sublattices. Since this study aims to demonstrate quantum oscillations in the spectral function, the strong finite-size effect in the real-space calculation must be suppressed. In the conventional periodic boundary conditions (PBC), the positions of the discrete momenta are almost concentric with the equal-energy contour around the Dirac point, causing sharp finite-size spikes in the DOS even in the presence of spectral broadening. To reduce this finite-size degeneracy, we use the shifted boundary condition (SBC) introduced in Kitaev's original paper \cite{Kitaev2006}, where the lattice is constructed on a torus with the basis $(L\ba_1, L\ba_2+\ba_1)$. In the following, we present the data from the real-space exact diagonalization of $L=122$ (29768 sites) systems on a torus with SBC and apply the Lorentzian broadening on the spectral data \footnote{Here we employ a relatively large Lorentzian broadening, $\gm = 0.1$, to fully suppress the finite-size-induced spectral spikes. In principle, a smaller broadening could be used for larger systems, either through exact diagonalization or through scalable spectral methods such as the kernel polynomial method (KPM), which avoids full diagonalization and enables calculations on substantially larger lattices. Physically, the phenomenological broadening may also mimic, at least in part, the finite-temperature smearing of the spectrum discussed in Ref.~\cite{Rossier2009,Knolle2020}.}. 
{Since the fluxes must appear in pairs on the torus, we prepare the flux sector with two isolated fluxes that are maximally separated on the lattice, that is, $\sim L/2$ unit cells for each direction. The on-site response is calculated from the right edge of one isolated flux plaquette and the sites to its right ( see insets of Fig.~\ref{Fig:STM_interval1}) up to a distance of $~\sim L/4$ unit cells. This setup minimizes the influence coming from the other isolated flux.} For each measuring site $j$, three different components $S^{xx}_{jj}$, $S^{yy}_{jj}$, and $S^{zz}_{jj}$ are calculated in the corresponding sector with the additional flux-pair insertion along the $x$-, $y$-, and $z$-bond, respectively.

\subsection{Results}

In Fig.~\ref{Fig:STM_interval1}, we present the results of different components of the flux-contrasted dynamical spin correlation functions
\begin{align} \label{eq:DeltaS}
\Dt S^{\al\al}_{jj} (\om') = \bar{S}^{\al\al}_{jj, \pi}(\om')-\bar{S}^{\al\al}_{jj, 0}(\om'),
\end{align}
where the subscripts $\pi$ and $0$ denote the system in the presence and absence of an isolated flux at the center. We shift the spectrum by the constant two-flux gap and define the frequency and the corresponding wavevector as
\begin{align}
\om' \equiv \om - \Dt_{2f}, \qquad k_{\om'} \equiv \frac{\sqrt{(\om')^2-\Dt^2}}{v_s}.
\end{align}
The bar symbol denotes a normalization $\bar{S}^{\al\al}_{jj} = S^{\al\al}_{jj}/\rho_0^{\rm lat.}$, where $\rho_0^{\rm lat.}$ is the Dirac linear density of states.
This normalization facilitates comparison with the low-energy effective-theory calculation in later sections. 
{The inset of Fig.~\ref{Fig:STM_interval1} shows the real-space setup, including the measuring site $j$ and the flux sector for calculating the matrix element in $S^{\al\al}_{jj,\pi}$ .} Note that we consider $r > 10$ responses away from the central flux, such that the interaction between the central flux (gray hexagon in the inset) and the tip fluxes (colored hexagons in the inset) is negligible. {This results in a spatially independent two-flux gap $\Dt_{2f} \approx 0.26$ in the spectrum, such that we can use the shifted quantities $\om'$ and $k_{\om'}$ in the data presentation.}

In Fig.~\ref{Fig:STM_interval1}, we show the tunneling spectra for a sequence of lattice sites to the right of an isolated $\pi$-flux excitation. On each measuring site, different components ($\al = x, y, z$) of the dynamical spin correlation are calculated under the flux sectors with corresponding flux-pair insertion (colored hexagons in the inset). In this specific orientation, the flux-pair insertion generated by $\sigma^z_j$ is identical for A- and B-sublattice sites, leading to equivalent response of $\Dt S^{zz}_{jj}$ shown in Fig.~\ref{Fig:STM_interval1}(c) and (f).
In contrast, $\Dt S^{xx}_{jj}$ for the A site and $\Dt S^{yy}_{jj}$ for the B site are related by the reflection symmetry of the flux-sector geometry (and so does $\Dt S^{xx}_{jj}$ for the B site and $\Dt S^{yy}_{jj}$ for the A site), such that the total response $\sum_{\al} \Dt S^{\al\al}_{jj}$ is sublattice-independent. {Note that the above discussion applies to the other orientations related by $C_6$ crystallographic rotation, and the behaviors of different spin components are permuted accordingly.}

When we compare the data for a fixed position $r$ and its neighboring cell $r+1$, a prominent phase shift of the oscillation is present, originating from the intervalley interference. This phase shift has a structure
\begin{align}
\phi_n = (\bK-\bK')\cdot\bR_n = \frac{2\pi}{3}n, \label{eq:Period3}
\end{align}
where $\bK$ and $\bK'$ are the two valley momenta and $\bR_n$ is the Bravais vector of the $n$-th unit cell. This indicates that, if we choose to present the data for every third unit cell, $\Dt r = 3$, instead of every unit cell, $\Dt r = 1$, this relative phase shift from the intervalley effect can be removed. In Fig.~\ref{Fig:STM_interval3}, we present the data under this special selection of sites. When plotting the flux-contrasted dynamical spin correlation as a function of $k_{\om} r$, as in Fig.~\ref{Fig:STM_interval3}(c-f), the intervalley interference is reduced and the oscillations for different $r$ are better aligned. This provides a cleaner presentation of the oscillation profile and also makes it comparable to the analytical results from the low-energy effective theory discussed in the next section.

\section{Effective low-energy model}\label{Sec:continuum_model}

In this section, we study the low-energy effective theory of the fermionic local density of states. 
We first map the problem to a
continuum Dirac Hamiltonian with a U(1) vector potential in the symmetric gauge, and use its analytical solutions to construct the retarded Green's function. The corresponding local density of states exhibits the Aharonov-Bohm oscillation in both frequency and tip-flux separation. By matching the spectral broadening and incorporating flux-pair insertion via T-matrix resummation, we show that the lattice and continuum calculations yield consistent flux-contrasted LDOS.

\subsection{Continuum model with a $\pi$-flux}\label{subsec:continnum_model_pi_flux}

In this section, it is convenient to introduce two independent replicas~\cite{Zuber1977,Halasz2016} of the Majorana hopping model, Eqs.~\eqref{eq:HMajo}, and combine the two replicated Majorana operators into a complex fermion on each site; see App.~\ref{app:EffectiveU1Theory} and\ref{app:GFComplex} for details. Note that this procedure involves an O(2) replica symmetry equivalent to particle number conservation of complex fermions. We exploit this symmetry and the fact that the on-site Green's function of interest is independent of static U(1) gauge transformations to incorporate the isolated $\pi$ flux of the Kitaev model as a U(1) vector potential in symmetric gauge. Focusing on low energies, we describe the system using a continuum Hamiltonian supplemented by a flux solenoid. In the symmetric gauge, we have the vector potential $\bA = \frac{\Phi}{2\pi r}\hat{\phi}$ and $\Phi$ is the magnetic flux. The low-energy Hamiltonian is
\begin{align}\label{eq:HCont}
\hat{\calH}_{\zt}(\bm{\Pi}) = v_s\left[\zt\sg_x\Pi_{x}+\sg_y\Pi_{y}\right]+\Dt\zt\sg_z,
\end{align}
where $\bm{\Pi} = \bp+\bA$ and $\zt = \pm 1$ labels the two valleys.
Technically, an isolated $\pi$-flux excitation on the hexagon of the lattice model also creates a local impurity potential which contributes to the intervalley scattering responsible for the period three phase shifts discussed around Eq.~\eqref{eq:Period3}. At the same time, it can be expected that Eq.~\eqref{eq:HCont} contains the main long-distance physics.

{The flux parameter is defined as $\xi \equiv \frac{\Phi}{2\pi}$. Unlike the true magnetic flux that can be tuned continuously, the flux in our system comes from the intrinsic $\mathbb{Z}_2$ excitation of the Kitaev spin liquid such that both $\Phi = \pm \pi$ correspond to the same $\mathbb{Z}_2$ sector. In the following, we use $\Phi = 0$ ($\xi = 0$) to denote the zero-flux case and $\Phi = \pi$ ($\xi = 1/2$) to denote the $\pi$-flux case.}

A similar Hamiltonian and its solution have been discussed in the context of a two-dimensional electron gas in the presence of extrinsic magnetic flux \cite{Slobodeniuk2010}, and we review this solution in App.~\ref{app:SolDirac}. In our case, we use the continuum model as the low-energy effective theory of charge-neutral fermionic excitation in the presence of an intrinsic $\pi$-flux excitation of the Kitaev spin liquid.

{Based on the exact eigenstates, we derive the retarded Green's function that is a $4\times 4$ matrix in the combined valley and sublattice spaces
\begin{align}\label{eq:GF_matrix}
\begin{split}
&\bG^R_{\Phi}(\br, \br'; \om) = \bg g_{\Phi,K}(\br,\br';\om) & 0 \\ 0 & g_{\Phi,K'}(\br,\br';\om) \ed,
\end{split}
\end{align}
where
\begin{align}
\begin{split}
&g_{\Phi,K}(\br,\br';\om) = \bg G^{R,AA}_{\Phi,K}(\br,\br';\om) & G^{R,AB}_{\Phi,K}(\br,\br';\om) \\[3pt]  G^{R,BA}_{\Phi,K}  (\br,\br';\om)& G^{R,BB}_{\Phi,K}(\br,\br';\om) \ed,\\
&g_{\Phi,K'}(\br,\br';\om) = \bg G^{R,AA}_{\Phi,K'}(\br,\br';\om) & G^{R,AB}_{\Phi,K'}(\br,\br';\om) \\[3pt] G^{R,BA}_{\Phi,K'}(\br,\br';\om) & G^{R,BB}_{\Phi,K'}(\br,\br';\om) \ed.
\end{split}
\end{align}

The detailed derivation and all the components of the retarded Green's function are given in App.~\ref{app:bareGF}.

For the later T-matrix treatment of the effective local potential induced by the flux-pair insertion (see Sec.~\ref{sec:T_matrix}, below), it will be convenient to perform a basis transformation of the valleys $
\ket{\pm} = (\ket{K}\pm\ket{K'})/\sqrt{2}$%simplifies the derivation:
\begin{align}
\begin{split}
\tilde{\bG}^R_{\Phi} (\br,\br';\om) &= \frac{1}{2}\bg \bm{1} & \bm{1} \\ \bm{1} & -\bm{1} \ed \bg g_{\Phi,K} & 0 \\ 0 & g_{\Phi,K'} \ed \bg \bm{1} & \bm{1} \\ \bm{1} & -\bm{1} \ed \\ &\equiv \bg g_{\Phi,+}(\br,\br';\om) & g_{\Phi,-}(\br,\br';\om) \\ g_{\Phi,-}(\br,\br';\om) & g_{\Phi,+}(\br,\br';\om) \ed,
\end{split}
\end{align}
where $g_{\Phi,\pm} = (g_{\Phi,K}\pm g_{\Phi,K'})/2$.

Since we are interested in the on-site Green's functions and the radial dependence of the LDOS, we can simplify the expression by setting $r = r'$ and $\phi = \phi' = 0$, and the resulting bare Green's functions in the zero-flux and $\pi$-flux cases have the structure:
\begin{align}\label{Eq:onsite_GF_matrix}
g_{\Phi,+} &= \bg \calI_{\Phi} & \calY_{\Phi} \\ -\calY_{\Phi} & \calI_{\Phi} \ed, \quad g_{\Phi,-} = \bg \calZ_{\Phi} & \calX_{\Phi} \\ \calX_{\Phi} & -\calZ_{\Phi} \ed,
\end{align}
where
\begin{subequations}
\begin{align}
&\calI_{\Phi} = -\frac{i\om}{4v_s^2}\left[S_{\Phi}(x)+\frac{1}{2}\left(1-\frac{\Dt}{\om}\right)\dt_{\Phi}(x)\right],\\
&\calZ_{\Phi} = -\frac{i\Dt}{4v_s^2}\left[S_{\Phi}(x)-\frac{1}{2}\left(\frac{\om}{\Dt}-1\right)\dt_{\Phi} (x)\right],\\
& \calY_{\Phi} = {\frac{x}{8 v_s r}\dt_{\Phi}(x)},\\
&\calX_{\Phi} = {\frac{-1}{4\pi v_s r}\left[2\Ld r + \frac{\pi x}{2}e^{-i 2x}\dt_{\Phi}(x)\right]}.
\end{align}
\end{subequations}

Here, $x = k_{\om} r$ is a dimensionless variable and $k_{\om} = \sqrt{\om^2-\Dt^2}/v_s$ is the frequency-dependent wavevector. The functions $S_\Phi(x), \delta_\Phi(x)$ are given by sums over products of Bessel functions of the first kind, $J_n(x)$, and Bessel functions of the second kind, $Y_n(x)$:
\begin{subequations}\label{eq:S_and_delta}
\begin{align}
    S_0(x) & = 1 + i \sum_{n= -\infty}^\infty J_n(x)Y_n(x), \\
    S_{\pi}(x) & = \frac{2}{\pi} \operatorname{Si}(2x)  + 2 i \sum_{n=0}^{\infty} J_{n + 1/2}(x)Y_{n + 1/2}(x),\\
\delta_0(x) & = 0,\\ 
\delta_{\pi}(x)  &= \frac{2}{\pi x} e^{i 2x},
\end{align}
\end{subequations}
where $\operatorname{Si}(2x)$ is the sine integral function.
Because parts of these sums are UV divergent, we introduce $\Ld \sim 1/a$ in $\calX_{\Phi}$ as the momentum cutoff for regularizing the infinite sum of partial waves, and $a$ is the lattice constant.

{Detailed derivations for the above components of the retarded Green's function are provided in App.~\ref{app:GF_components}.}

\subsection{LDOS from the bare Green's function}

\begin{figure}
\includegraphics[width=1.0\columnwidth]{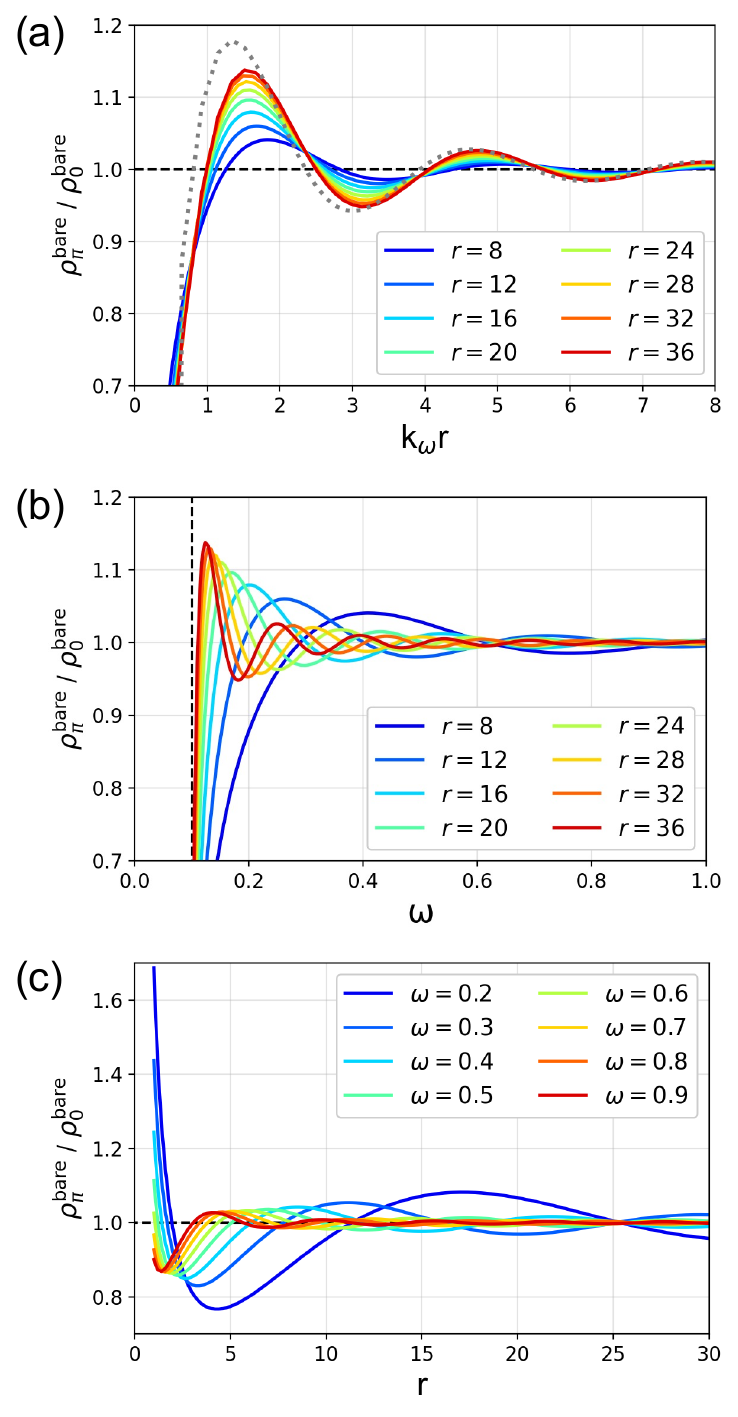}
     \caption{\label{fig:bare_LDOS} The bare LDOS of the continuum model in the presence of a $\pi$ flux, calculated by Eq.~(\ref{Eq:bare_LDOS}). The linear LDOS is defined as a background reference $\rho_{0}^{\rm bare}$. In (a), each curve is evaluated at fixed $r$ and $k_{\om} = \sqrt{\om^2-\Dt^2}/v_s$ is $\om$-dependent. The horizontal black dashed line denotes the reference value for the zero-flux case. The gray dotted line shows the approximate form in the far-field limit for $r = 36$, calculated by Eq.~(\ref{Eq:bare_LDOS_approximated}). In (b), the vertical black dashed line denotes the mass gap $\Dt = 0.1$, which is the constant used in all three subfigures. }
\end{figure}

The LDOS on the A sublattice is defined as
\begin{align}
\begin{split}
&\rho^{\rm bare}_{\Phi, A}(r, \om) = -\frac{1}{\pi}\operatorname{Im}\left[2g_{\Phi,+}(\br,\br;\om)\right]_{AA} = -\frac{2}{\pi}\operatorname{Im}\calI_{\Phi}, \label{eq:rhobare}
\end{split}
\end{align}
which is identical for both sublattices. Therefore, we drop the sublattice index in the LDOS below. In the zero-flux case,
\begin{align}
\begin{split}
&\rho^{\rm bare}_0 (r,\om) = 
\frac{\om}{2\pi v_s^2},
\end{split}
\end{align}
while in the $\pi$-flux case 
\begin{align}
\begin{split}\label{Eq:bare_LDOS}
\rho^{\rm bare}_{\pi}(r, \om)
&=\frac{\om}{2\pi v_s^2}\left[\frac{2\operatorname{Si}(2x)}{\pi}+\left(1-\frac{\Dt}{\om}\right)\frac{\cos(2x)}{\pi x}\right].
\end{split}
\end{align}

One useful quantity for visualizing the quantum oscillation induced by the presence of the $\pi$-flux is the flux-contrasted LDOS:
\begin{align}
\Dt \rho^{\rm bare} = \rho^{\rm bare}_{\pi} - \rho^{\rm bare}_{0} = -\frac{1}{\pi}\operatorname{Im}(2\Dt\calI),
\end{align}
where $\Dt\calI = \calI_{\pi}-\calI_0$.
In the far-field limit, $x = k_{\om} r\gg1$, the asymptotic form of the sine integral is $\mathrm{Si}(2x)\simeq \frac{\pi}{2}-\frac{\cos(2x)}{2x}$, such that
\begin{align}\label{Eq:bare_LDOS_approximated}
\rho^{\rm bare}_{\pi}(r,\om) \simeq \frac{\om}{2\pi v_s^2}\left[1-\frac{\Dt}{\om}\left(\frac{\cos2x}{\pi x}\right)\right]
\end{align}
and
\begin{align}
\Dt\rho^{\mathrm{bare}}(r,\om) \simeq -\frac{1}{2\pi^2}\frac{\Dt\cos(2x)}{v_s^2 x}. \label{eq:deltarho}
\end{align}
In Fig.~\ref{fig:bare_LDOS}, we plot the bare LDOS in the presence of a $\pi$ flux as a function of $k_{\om}r$, $\om$, and $r$. In Fig.~\ref{fig:bare_LDOS}(a) and (b), each curve corresponds to a fixed tip-flux separation $r$, as in the STM measurement performed at fixed tip position while varying the bias voltage. On the other hand, in Fig.~\ref{fig:bare_LDOS}(c), each curve is evaluated at fixed $\om$ and shows the oscillation as a function of $r$, corresponding to a scan of the tip position at fixed bias.

We note that Eq.~\eqref{eq:deltarho} in particular holds near the bottom of the band. For this limit of parabolic dispersion, we compare Eq.~\eqref{eq:deltarho} with the result for Friedel oscillations, for which in the limit of a weak point-like potential $U_p \ll v_s^2/\Delta$,
\begin{align}
\Delta \rho^{\rm bare} \simeq -\frac{U_p \Delta}{2\pi^2 v_s^2} \frac{\Delta \cos(2x)}{v_s^2 x}.
\end{align}
While the functional dependence is the same, the small dimensionless prefactor is fundamentally distinct from the order-one amplitude of oscillations in  Eq.~\eqref{eq:deltarho}. In the reverse limit $U_p \gtrsim v_s^2/\Delta$, the full T-matrix resummation generically leads to a phase shift of the Friedel oscillations as compared with Eq.~\eqref{eq:deltarho}.

\subsection{Spectral broadening}

To compare with the lattice-model exact diagonalization results, we apply the same broadening factor to the retarded Green's function. This can be done by replacing the variables:
\begin{align}
\om \to \Om \equiv \om+i\gm, \quad x \to z \equiv \frac{\sqrt{\Om^2-\Dt^2}}{v_s}\, r,
\end{align}
where the imaginary frequency $\gm$ is the broadening factor, and we focus on the branch with $\operatorname{Im}\sqrt{\Om^2-\Dt^2} > 0$.
The broadened $\calI_0^{\gm}$ can be derived by using the translational invariance
\begin{align}
\begin{split}
&G^{R,\gm}_{0} (\bk,\Om) = \frac{\Om+v_s(k_x\sg_x+k_y\sg_y)+\Dt\sg_z}{\Om^2-\Dt^2-v_s^2 k^2},\\
&g_{0}^{\gm}(\Om) = \int_{|\bk|< \Ld}\frac{\mathrm{d}k^2}{(2\pi)^2}G^{R,\gm}_{0}(\bk, \Om).
\end{split}
\end{align}
Since we only need the $g_{0,+}^{\gm} = \calI_0^{\gm} \sg_0$ channel for the LDOS, and $k_x$ and $k_y$ terms are integrated over symmetric space, we have
\begin{align}
\begin{split}
\calI_0^{\gm}(\Om) &= \Om\int_{|\bk|<\Ld} \frac{\mathrm{d}k^2}{(2\pi)^2}\frac{1}{\Om^2-\Dt^2-v_s^2k^2}\\
&= \frac{\Om}{4\pi v_s^2}\ln\left[\frac{\Om^2-\Dt^2}{\Om^2-\Dt^2-v_s^2\Ld^2}\right],
\end{split}
\end{align}
The LDOS, which is proportional to the imaginary part of $\calI_0^{\gm}$, contains the momentum cutoff $\Ld$, except for $\gm \to 0^{+}$ 
\begin{align}
\calI_0 (\om)= \frac{\om}{4\pi v_s^2}\ln\left[\frac{\om^2-\Dt^2}{v_s^2\Ld^2-(\om^2-\Dt^2)}\right] -i\frac{\om}{4v_s^2},
\end{align}
leading to the standard LDOS without the spectral broadening. 

To calculate the flux-contrasted LDOS, we apply the same substitution for $\Dt \calI^{\gm}$ and get
\begin{align}\label{Eq:bare_flux_constrasted_LDOS}
\begin{split}
\Dt \rho^{\rm bare}_{\gm}\,(r,\Om) &= \frac{1}{2\pi v_s^2}\operatorname{Re}\left[\Om \,S_{\pi}(z)-\Om\, S_0 (z)\right]\\
&\qquad +\frac{1}{4\pi v_s^2}\operatorname{Re}\left[\left(\Om-\Dt\right)\,\dt_{\pi}(z)\right].
\end{split}
\end{align}
Even though both $S_{\pi}$ and $S_{0}$ are $\Ld$-dependent, their difference removes the regularized divergence and gives $\Ld$-independent $\Dt\rho^{\rm bare}_{\gm}$. 

In Fig.~\ref{fig:bare_LDOS_compare}, we compare the flux-contrasted LDOS between the continuum model, Eq.~(\ref{Eq:bare_flux_constrasted_LDOS}), and the Kitaev lattice model calculation. In the lattice calculation, we have an isolated flux at the center of the lattice, and the distance $r$ is estimated as the number of unit cells from the central plaquette. Note that we imagine that the $\pi$ flux is located at the center of the central plaquette, such that the positions of the measuring sites have an additional shift of $0.5$ unit-cell distance. 

In this comparison, we intend to focus on the flux-induced oscillation in the LDOS without the additional flux-pair insertion required in the spin correlation function of the Kitaev model. Specifically, we calculate the matrix element in Eq.~(\ref{Eq:onsite_spin_correlation}) in the flux sector with the central isolated flux, but without the additional flux pair around the measuring site. 

Even for $L=122$, the lattice spectrum contains visible finite-size spikes that can contaminate the Aharonov-Bohm oscillation. We therefore use a relatively large Lorentzian broadening, $\gm = 0.1$, to smooth these unphysical spikes. When using the same $\gm$ in the continuum model, it introduces an approximate damping term $e^{-2k_I r}$ where $k_I$ is the imaginary part of the wavevector generated by the broadening, $\Om^2 = v_s^2(k_{\om}+ik_I)^2+\Dt^2$. This damping term suppresses the LDOS magnitude for larger $r$ in Fig.~\ref{fig:bare_LDOS_compare}, in contrast to the zero-broadening result in Fig.~\ref{fig:bare_LDOS}.

\begin{figure}
\includegraphics[width=1.0\columnwidth]{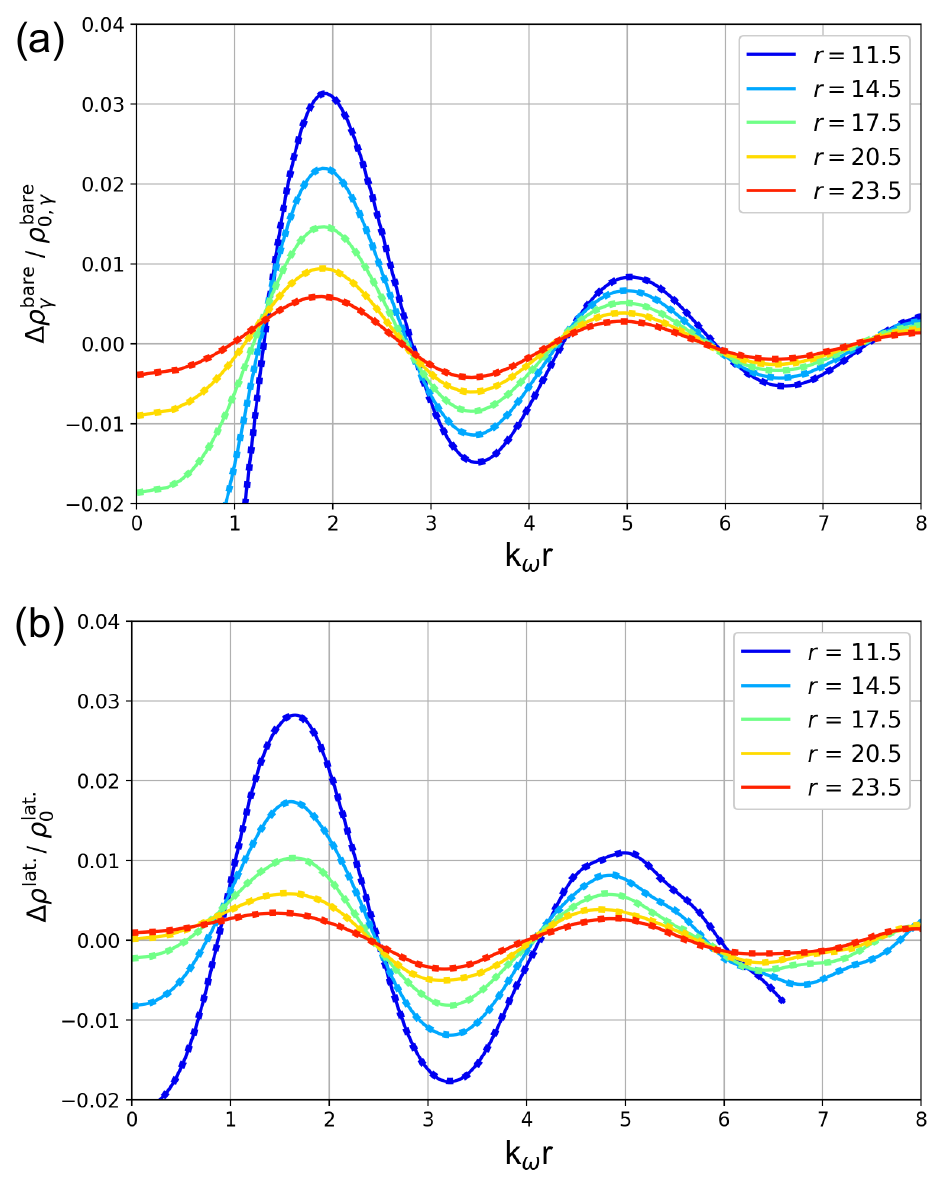}
     \caption{\label{fig:bare_LDOS_compare} {The flux-contrasted LDOS for (a) the continuum model, and (b) the lattice model ($L=122$) without the flux-pair insertion.} The continuum model result is calculated by Eq.~(\ref{Eq:bare_flux_constrasted_LDOS}) and normalized by $\rho_{0,\gm}^{\mathrm{bare}} = -\frac{1}{\pi}\operatorname{Im}[2\calI_0^{\gm}]$, where the broadening factor $\gm$ is chosen to be the same as the spectral broadening of the lattice result. The solid (dotted) line represents the flux-contrasted LDOS for A (B) sublattice sites.}
\end{figure}

\subsection{Flux-pair insertion and effective local potential}\label{sec:T_matrix}

In Sec.~\ref{Sec:dynamical_spin_correlation}, we showed that the dynamical on-site spin correlation function of the Kitaev spin liquid can be calculated in the two-flux excitation sector in the adiabatic approximation, where the additional flux-pair is inserted by flipping the local variable $u_{\langle jk\rangle_{\al}}$ at the measuring site. Therefore, in the continuum description, this effect can be incorporated as an effective local potential, whose explicit expression is derived in App.~\ref{app:effective_potential}.

\begin{figure}
\includegraphics[width=1.0\columnwidth]{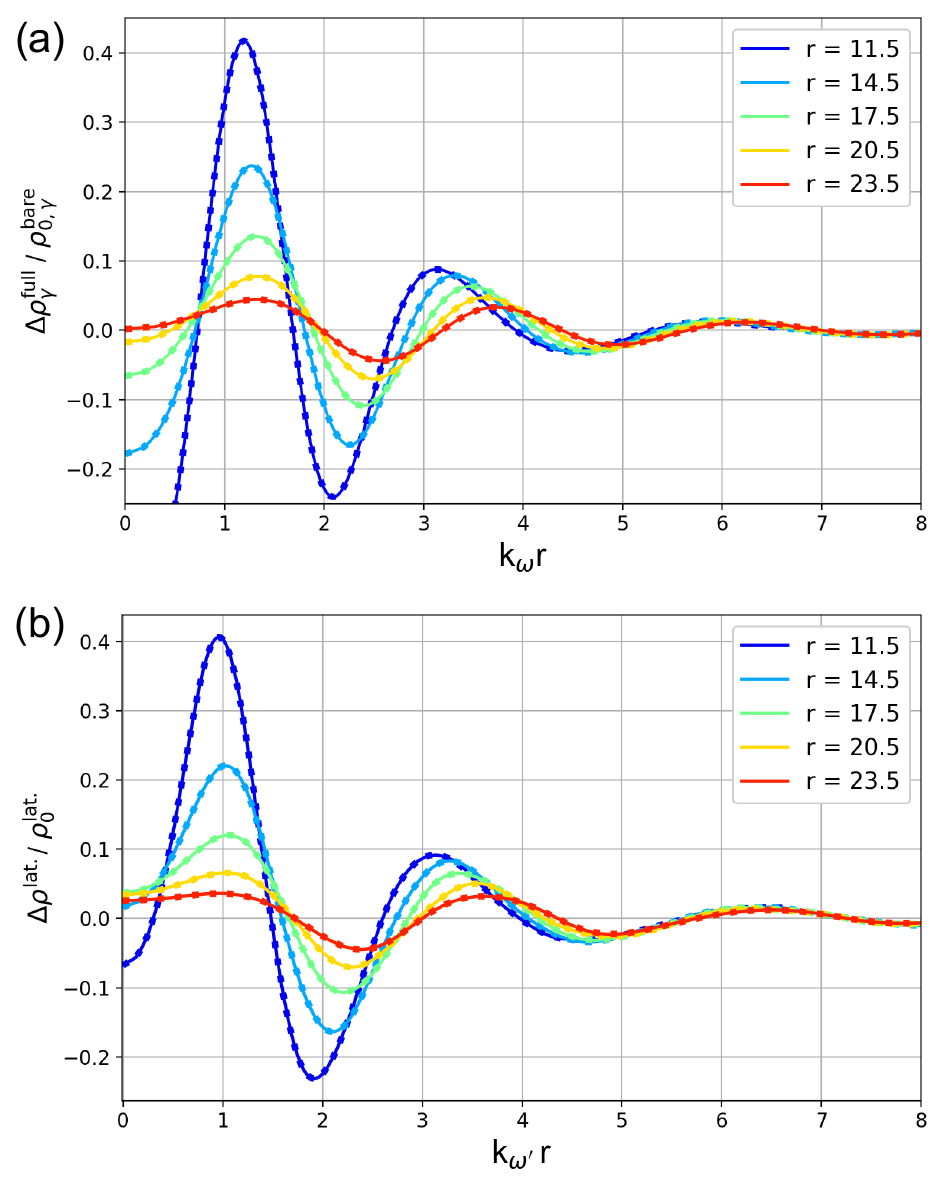}
     \caption{\label{fig:full_LDOS_compare} {The flux-contrasted LDOS for the (a) continuum model with effective local potential, and (b) lattice model ($L=122$) with the z-bond flux-pair insertion.} The continuum model result is calculated by Eq.~(\ref{Eq:tip_flux_constrasted_LDOS}) and normalized by $\rho_{0,\gm}^{\mathrm{bare}} = -\frac{1}{\pi}\operatorname{Im}[2\calI_0^{\gm}]$, where the broadening factor $\gm$ is chosen to be the same as the spectral broadening of the lattice result. The solid (dotted) line represents the flux-contrasted LDOS for A (B) sublattice sites.}
\end{figure}

Since the local potential is point-like at the measuring position, the full Green's function can be written as
\begin{align}\label{eq:full_GF}
 \mathbb{G}^R_{\Phi}(\br,\br; \om) = \left[\left(\bG^{R}_{\Phi}\right)^{-1}-\bm V\right]^{-1},
\end{align}
where $\bm V = U(\bm{1}+\tau_x)\otimes\sg_x$ and $U = -4J_z$. 

In the $\ket{\pm}$ basis of the valley wavefunctions, the potential is present only in the upper-left block
\begin{align}\label{eq:flux_pair_potential_matrix}
\tilde{\bm V} = \bg 2U\sg_x & 0 \\ 0 & 0 \ed,
\end{align}
such that the problem reduces to equations involving $2\times 2$ matrices in the sublattice space. The Dyson equation and the T-matrix are
\begin{align}
\begin{split}
&\tilde{\mathbb{G}}^R_{\Phi} = \tilde{\bG}^R_{\Phi}+\tilde{\bG}^R_{\Phi}\tilde{\bm T}_{\Phi}\tilde{\bG}^R_{\Phi}, \\
&\tilde{\bm T}_{\Phi} = \tilde{\bm V}(\bm 1-\tilde{\bG}^R_{\Phi}\tilde{\bm V})^{-1} \equiv \bg t_{\Phi,+} & 0 \\ 0 & 0  \ed,
\end{split}
\end{align}
where $t_{\Phi,+} = \tilde{u}(1-g_{\Phi,+}\tilde{u})^{-1}$ and $\tilde{u} = 2U\sg_x$.

For the full on-site LDOS, it contains only the $\tilde{\mathbb{G}}^{R}_{\Phi,++} = g^{\phantom{R}}_{\Phi,+}+g^{\phantom{R}}_{\Phi,+}t^{\phantom{R}}_{\Phi,+}g^{\phantom{R}}_{\Phi,+}$ component:
\begin{align}
\begin{split}
\rho^{\rm full}_{\Phi} &= -\frac{1}{\pi}\rm{Im}\rm{Tr}_{\sg}\left[g^{\phantom{R}}_{\Phi,+}+g_{\Phi,+}^2 t^{\phantom{R}}_{\Phi,+}\right].
\end{split}
\end{align}

Therefore, the flux-contrasted LDOS is
\begin{align}
\begin{split}
&\Delta \rho^{\rm full} = \rho^{\rm full}_{\pi} - \rho^{\rm full}_0= -\frac{1}{\pi}\rm{Im}\rm{Tr}_{\sg}\left[(g^{\phantom{R}}_{\pi,+}-g^{\phantom{R}}_{0,+})\right.\\
&\qquad\qquad\qquad\qquad\qquad\left.+g_{\pi,+}^2t^{\phantom{R}}_{\pi,+}-g_{0,+}^2t^{\phantom{R}}_{0,+}\right].
\end{split}
\end{align}

From the previous definitions, we know that
\begin{align}
g_{0,+}^2 = \calI_0^2 \sg^{\phantom{R}}_0, \quad g_{\pi,+}^2 = (\calI_{\pi}^2-\calY_{\pi}^2)\sg^{\phantom{R}}_0 + 2\calI_{\pi}\calY_{\pi}(i\sg_y),
\end{align}
such that the T-matrix is
\begin{align}
\begin{split}
&t_{\Phi,+} = \frac{2U}{\mathcal{D}_{\Phi,+}}\bg 2U\calI_{\Phi} & 1-2U\calY_{\Phi} \\ 1+2U\calY_{\Phi} & 2U\calI_{\Phi} \ed, \\
&\mathcal{D}_{\Phi,+} = 1-4U^2\left(\calI_{\Phi}^2+\calY_{\Phi}^2\right).
\end{split}
\end{align}

To calculate the LDOS, we first derive
\begin{align}
\begin{split}
&\mathrm{Tr}_{\sg}\left\{g_{\Phi,+}^2\left[\calI_{\Phi}\sg_0+\sg_x-i\calY_{\Phi}\sg_y\right]\right\} =2\calI_{\Phi}\left(\calI_{\Phi}^2+\calY_{\Phi}^2\right),
\end{split}
\end{align}
%and define the quantity
%\begin{align}
%\calT_{\Phi} \equiv \frac{16U^2\calI_{\Phi}\left(\calI_{\Phi}^2+\calY_{\Phi}^2\right)}{1-4U^2\left(\calI_{\Phi}^2+\calY_{\Phi}^2\right)}.
%\end{align}
and for the flux-contrasted LDOS with the local potential, we obtain
\begin{align}
\begin{split}
\rho^{\rm full}_{\Phi} &= -\frac{1}{\pi}\mathrm{Im}\left[\frac{2\calI_{\Phi}}{1-4U^2\left(\calI_{\Phi}^2+\calY_{\Phi}^2\right)}\right],
\end{split}
\end{align}
and
\begin{align}
\Dt\rho^{\rm full} = -\frac{1}{\pi}\mathrm{Im}\left[\frac{2\calI_{\pi}}{1-4U^2\left(\calI_{\pi}^2+\calY_{\pi}^2\right)}-\frac{2\calI_0}{1-4U^2\calI_0^2}\right].
\end{align}

For the broadened version of $\Dt\rho^{\rm full}$, we again introduce $\om\to\Om = \om+i\gm$ and $x\to z$ into $\calI_{\Phi}$ and $\calY_{\pi}$. We can calculate $\calI_{\pi}^{\gm} = \calI_{0}^{\gm} + \Dt\calI^{\gm}$ and
\begin{align}
\calY_{\pi}^{\gm}(z) = \frac{1}{4\pi v r}e^{i2z}.
\end{align}

Therefore, we obtain the expression for the flux-contrasted on-site LDOS with the local potential and finite broadening:
\begin{align}\label{Eq:tip_flux_constrasted_LDOS}
\begin{split}
&\Dt\rho^{\mathrm{full}}_{\gm} = -\frac{1}{\pi}\mathrm{Im}\left[\frac{2\calI^{\gm}_{\pi}}{1-4U^2\left[(\calI_{\pi}^{\gm})^2+(\calY_{\pi}^{\gm})^2\right]}-\frac{2\calI^{\gm}_0}{1-4U^2(\calI_0^{\gm})^2}\right],
\end{split}
\end{align}
which contains the parameters $U$ and $\Ld$.

In Fig.~\ref{fig:full_LDOS_compare}, {we compare the flux-contrasted LDOS between the continuum model with an effective local potential using the T-matrix approach and the lattice Kitaev model with the local flux-pair insertion.} Note that in this case, the data from the lattice calculation are indeed the dynamical on-site spin correlation function in the adiabatic approximation, Eq.~(\ref{Eq:onsite_spin_correlation}), which contributes to the inelastic STM signal. One important observation is that the overall intensity in Fig.~\ref{fig:full_LDOS_compare} is much larger compared with the case without the local potential (flux-pair insertion) in Fig.~\ref{fig:bare_LDOS_compare}. This enhancement has the same origin as the resonance peak in the dynamical structure factor of the Kitaev model, which is strikingly different from the linear fermionic DOS \cite{Knolle2014, Knolle2015}. {Therefore, the presence of the local potential, or equivalently, the flux-pair insertion, greatly enhances the intensity of its LDOS oscillation at low energies in the Kitaev spin liquid compared with a conventional electronic Dirac system with a linear dispersion.}

\section{Conclusion}

In this work, we have established a local tunneling-spectroscopy
signature of an isolated flux excitation in the non-Abelian Kitaev
spin liquid.  In the charge-neutral inelastic cotunneling setup
considered here, the STM-IETS signal directly probes the on-site dynamical
spin correlation function of the Kitaev spin-liquid layer. By comparing the
response in the presence and absence of an isolated $\pi$ flux, we find oscillations both as a function of frequency and tip-flux separation. These oscillations can be quantitatively distinguished from fluxless Friedel-like oscillations by means of their amplitude and phase shift. The microscopic Kitaev-model calculation and the low-energy Dirac theory yield
consistent oscillatory structures controlled by $k_{\om}r$, while the lattice calculation additionally resolves the short-distance intervalley modulations that dress the long-wavelength radial envelope.  The agreement between large-scale lattice calculation and continuum theory identifies the
observed signal as an intrinsic flux-induced interference effect,
rather than a finite-size feature. 

Conceptually, our result extends the Aharonov-Bohm effect from its
conventional electronic setting to a fractionalized and electrically
neutral quantum system.  In an electronic system, the Aharonov-Bohm
phase is acquired by a charged particle propagating around an external
electromagnetic flux.  Here, the propagating object is a charge-neutral
Majorana excitation, while the enclosed flux is an intrinsic
$\mathbb{Z}_{2}$ excitation of the Kitaev spin liquid.  The vector
potential introduced in the continuum theory should therefore be
understood as an effective representation of the $\pi$ holonomy, rather
than a physical electromagnetic coupling to Majorana fermions.
The fact that the same oscillation is obtained directly in the lattice calculation of the microscopic Kitaev model confirms this interpretation.
Equivalently, closed Majorana trajectories that wind an odd number of
times around the flux acquire an additional minus sign.  In the
Ising-anyon description, this sign is the mutual full-braiding phase
$-1$ between the itinerant Majorana fermion $\psi$ and
the flux excitation $\sigma$.  The resulting oscillation is therefore
a braiding-related interference signature of the Ising topological
order, although it does not by itself form a complete
measurement of non-Abelian exchange operations or fusion-channel
dynamics. We remark in passing that the mutual braiding phase between the fermion and the flux is always nontrivial and equal to $-1$ in every topological phase accessible in Kitaev-like models distinguished by the Majorana-Chern number modulo $16$. While we concentrated on Ising topological order, similar results in STM-IETS are expected for other phases as well.

An essential ingredient of the physical STM-IETS response is that a
local spin operator in the Kitaev model does not probe a bare Majorana
propagator.  It inserts a flux pair at the tunneling position and thus produces a local quantum quench for the matter Majorana fermions. In the microscopic model, this is incorporated through the adiabatic approximation of the dynamical spin correlations. In the continuum model, we introduce a local-quench potential treated by the T-matrix formalism and show that the resulting LDOS is also consistent with the lattice calculation. Importantly, the Aharonov-Bohm oscillation is already present in the pure isolated-flux background,
before this local quench is introduced.  The STM-IETS measurement-induced flux
pair therefore does not generate the oscillation or suppress its
visibility.  Instead, near the resonance of the quenched Majorana
response, it substantially enhances the spectral intensity along with 
the oscillatory pattern.  The flux-pair insertion in the spin correlation function of the Kitaev model, which might appear
to complicate the detection of the underlying flux at first glance, thus actually improves the spectroscopic visibility of the effect. 

Our result also complements previous tunneling proposals for vacancy
defects in Kitaev spin liquids \cite{takahashi2023nonlocal,Kao2024PRL,Kao2024PRB, Li2026_planar}. Vacancy defects are interesting in the Kitaev spin liquid because they can trap $\pi$ fluxes in the ground state \cite{Willans2010,Willans2011,Kao2021vacancy, Kao2021localization,Vitor2022,Xiao2025}. Moreover, each vacancy introduces three dangling spin operators excluded from forming conserved plaquette-flux operators, such that their dynamical correlation does not include the local quantum quench and the two-flux excitation gap. Therefore, the STM-IETS signal right on top of the vacancy position is dominated by these dangling-spin correlations and reveals vacancy-induced Majorana modes near zero frequency. The present work addresses a distinct question: whether a preexisting $\pi$ flux can also be detected through the ordinary dynamical correlations of the bulk spins, for which the associated flux-pair insertion is unavoidable. Our results reveal the Aharonov-Bohm oscillation induced by a fractionalized flux excitation, and thus provide an affirmative answer to this question. Notably, this signature occurs in the finite-frequency bulk Majorana continuum and does not require a sharply resolved zero mode or a near-zero-energy resonance. In this sense, the Aharonov-Bohm oscillation, which encodes the mutual braiding phase effect, supports a complementary finite-energy diagnostic of the fractionalized flux excitation.

\begin{acknowledgements}
Both authors would like to thank Nandini Trivedi, Penghao Zhu, and Ryan Buechele for the helpful discussions during the development of the project. W.-H.K. would like to thank Natalia Perkins and G\'abor Hal\'asz for their prior collaboration on this topic.
Support for this research was provided by the Office of the Vice Chancellor for Research and Graduate Education at the University of Wisconsin–Madison with funding from the Wisconsin Alumni Research Foundation (W.-H.K. and E.J.K.).
\end{acknowledgements}

\appendix

\section{Tunneling spectroscopy of Kitaev spin liquids}
\label{app:tunneling}

This appendix contains supporting information for Sec.~\ref{sec:tunneling}. It is conceptually based on earlier works \cite{Konig2020, Knolle2020, Udagawa2021, Bauer2023, takahashi2023nonlocal, Kao2024PRL, Kao2024PRB, Zhang2025_PRB, Zhang2025_npj} and included to make this paper self-contained and to establish the notation used in the main text.

The total Hamiltonian of the setup (Fig.~\ref{Fig:schematic_setup}) can be broken down into
\begin{align}
\calH_{\rm IETS} = \calH_{\rm ele} + \calH_{\rm Kitaev} + \calH_{\rm tun} \equiv \calH_0 + \calH_{\rm tun},
\end{align}
where $\calH_{\rm ele}$ describes electrons in the metallic leads, $\calH_{\rm Kitaev}$ is the Kitaev spin-liquid Hamiltonian in Eq.~(\ref{Eq:Kitaev_Hamiltonian}), and $\calH_{\rm tun}$ is the tunneling Hamiltonian of the electrons in the presence of the spin-liquid layer. The first term contains two species of free electrons in the substrate and the tip, respectively,
\begin{align}
\calH_{\rm sub} =  \sum_{\mathbf{k}s} \varepsilon^{\phantom{\dg}}_{\mathbf{k}} \check{c}^{\dg}_{\mathbf{k}s} \check{c}^{\phantom{\dg}}_{\mathbf{k}s}, \qquad \mathcal{H}_{\mathrm{tip}} = \sum_{\mathbf{p}s'} \varepsilon^{\phantom{\dg}}_{\mathbf{p}} \hat{c}^{\dg}_{\mathbf{p}s'} \hat{c}^{\phantom{\dg}}_{\mathbf{p}s'}.
\end{align}
Note that $\hat{c}_{\bp s'}$ and $\check{c}_{\bk s}$ describe the spinful fermion operators and should not be confused with the $c$-Majorana operator used in the previous section. The general tunneling Hamiltonian can be written as
\begin{align}
\begin{split}
&\mathcal{H}_{\mathrm{tun}} = \sum_{\bk\bp}\sum_{ss'} T^{\phantom{\dg}}_{ss'} (\mathbf{r}) \, \check{c}^{\dg}_{\bk s} \hat{c}^{\phantom{\dg}}_{\bp s'} + \mathrm{H.c.},\\
&T^{\phantom{\dg}}_{ss'} (\mathbf{r}) = \sum_{j} \left[ T^{0}_{ss'} (\mathbf{r} - \mathbf{r}_j) + \vec{\sg}_j \cdot \vec {T}_{ss'} (\mathbf{r} - \mathbf{r}_j) \right],
\end{split}
\end{align}
where $s$ and $s'$ denote the spin indices of the electrons and $\vec{\sg}_j$ is the spin operator of the Kitaev spin-liquid layer. For simplicity, we assume that the tip is positioned directly above a given lattice site of the Kitaev layer, and the tunneling electron can only couple to the spin of that site,
\begin{align}
T_{ss'}(\br) \sim T_{ss'}(\br_j) \equiv t_0 \dt_{ss'} + t_1(\vec{\sg}_j\cdot \vec{\tau}_{ss'}),
\end{align}
where $t_0$ and $t_1$ are coefficients that depend on the microscopic details, and $\tau^{\al}$ are the Pauli matrices with $\al = x,y,z$.

Before and after the electron tunnels from the tip to the substrate, the initial and final many-body state of the total system can be written as a product state of different components
\begin{align}
\begin{split}
&\ket{\Psi} = \ket{\phi}\otimes \ket{\chi}\otimes\ket{n},\quad \ket{\Psi'} = \ket{\phi'}\otimes \ket{\chi'}\otimes\ket{m},
\end{split}
\end{align}
where $\ket{\phi'}$ has one less electron than $\ket{\phi}$, and $\ket{\chi'}$ has one more electron than $\ket{\chi}$. Assuming that the chemical potential of the tip is higher than that of the substrate by $eV$, the grand canonical energy difference is
\begin{align}
\begin{split}
E_{\Psi'}-E_{\Psi} &= (\varepsilon_{\bk}+\mu_s)-(\varepsilon_{\bp}+\mu_t) + E_{m}-E_{n}\\
&= \om_{mn} +  \varepsilon_{\bk}-\varepsilon_{\bp} -eV,
\end{split}
\end{align}
where $\om_{mn} = E_m - E_n$ and $\mu_t - \mu_s = eV$.
Therefore, we can derive the matrix element of the perturbation Hamiltonian
\begin{align}
\bra{\Psi'}\calH_{\rm tun}(t)\ket{\Psi} = e^{i(E_{\Psi'}-E_{\Psi})t}\bra{\Psi'}T_{ss'}(\br_j)\check{c}^{\dg}_{\bk s} \hat{c}^{\phantom{\dg}}_{\bp s'}\ket{\Psi}.
\end{align}

The transition probability is obtained from Fermi's golden rule as
\begin{align}
\begin{split}
\Gamma_{t\to s} = 2\pi\sum_{\Psi\Psi'}\sum_{\bk\bp}\sum_{ss'} & P_{\Psi} |\bra{\Psi'}T_{ss'}(\br_j)\check{c}_{\bk s}^{\dg} \hat{c}_{\bp s'}\ket{\Psi}|^2\\
& \times\dt(\om_{mn} +  \varepsilon_{\bk}-\varepsilon_{\bp} -eV),
\end{split}
\end{align}
where $P_{\Psi} = e^{-\bt E_{\Psi}}/Z_{\Psi}$ is the Boltzmann weight. The metallic part of the matrix element can be evaluated
\begin{align}
\sum_{\phi'\chi'} |\bra{\phi'\chi'}\check{c}_{\bk s}^{\dg}\hat{c}_{\bp s'}\ket{\phi\chi}|^2 = n_F(\varepsilon_{\bp})\left[1-n_F(\varepsilon_{\bk})\right],
\end{align}
where $n_F(\varepsilon)$ is the Fermi-Dirac function. At zero temperature, this becomes $n_F(\varepsilon_{\bp})\left[1-n_F(\varepsilon_{\bk})\right] \to \Theta(-\varepsilon_{\bp})\Theta(\varepsilon_{\bk})$ where $\Theta (\varepsilon)$ is the unit step function. Note that the delta function in the transition probability requires $\varepsilon_{\bk}-\varepsilon_{\bp} + E_{m}-E_{n} -eV = 0$, and at $T = 0$ we require $\varepsilon_{\bp} < 0$ and $\varepsilon_{\bk} >0$, such that $eV > (E_{m}-E_{n})$. This condition implies that the bias energy needs to exceed the magnetic excitation energy in order to see the tunneling current. At $T = 0$, the reverse process, $\Gamma_{s\to t}$ is blocked by a similar argument. One can further replace the momentum sums by integrals over the DOS and employ the flat-DOS approximation for the tip and the substrate $\sum_{\bk\bp} \sim D^{\rm tip}_{s'}D^{\rm sub}_s \int \mathrm{d}\varepsilon_{\bp}\mathrm{d}\varepsilon_{\bk}$, such that the integral can be evaluated
\begin{align}
\begin{split}
\int_0^{\infty}\mathrm{d}\varepsilon_{\bk}\int_{-\infty}^0\mathrm{d}\varepsilon_{\bp}&\dt(\om_{mn}+\varepsilon_{\bk}-\varepsilon_{\bp}-eV)\\
&= (eV-\om_{mn})\Theta(eV-\om_{mn}).
\end{split}
\end{align}

The tunneling current becomes
\begin{align}
\begin{split}
I = 2\pi e \sum_{mn}\sum_{ss'} &D^{\rm tip}_{s'}D^{\rm sub}_{s}|\bra{m}T_{ss'}(\br_j)\ket{n}|^2\\
&\times(eV-\om_{mn})\Theta(eV-\om_{mn}),
\end{split}
\end{align}
where the matrix element for the spin-liquid layer can be decomposed into three contributions
\begin{widetext}
\begin{align}
|\bra{m}T_{ss'}(\br_j)\ket{n}|^2  = t_0^2 \dt_{mn}\dt_{ss'} + 2t_0t_1 \dt_{mn}\tau^z_{ss'}\dt_{ss'}\langle \sg_j^z\rangle + t_1^2 \sum_{\al\bt} \tau^{\al}_{s's}\tau^{\bt}_{ss'} \bra{n}\sg^{\al}_j\ket{m}\bra{m}\sg^{\bt}_{j}\ket{n}.
\end{align}
Note that the first (second) term corresponds to the spin-independent (spin-dependent) elastic current. Because of the Kronecker $\delta_{nm}$ we recover $I = G V$, Eq.~\eqref{eq:IelMaintext}, provided that the density of states is nearly constant at the Fermi energy of the tip and substrate. In the STM-IETS signal, we focus on the third term, which corresponds to the spin-dependent inelastic current. 
At zero temperature, the initial state of the magnetic layer is the ground state of the Kitaev spin liquid, $\ket{n} \to \ket{0}$, and the corresponding spectral function $S^{\al\bt}_{jj}(\om)$ is nonzero only for $\al = \bt$. Therefore, the inelastic tunneling current is
\begin{align}
I_{\rm inel} = 2\pi e t_1^2\sum_{ss'}D^{\rm tip}_{s'}D^{\rm sub}_{s}\sum_{\al}\tau^{\al}_{ss'}\tau^{\al}_{s's}\int_0^{eV}\mathrm{d}\om (eV-\om)S^{\al\al}_{jj}(\om), \quad S^{\al\al}_{jj}(\om) =  \sum_{m}\bra{0}\sg^{\al}_j\ket{m}\bra{m}\sg^{\al}_j\ket{0}\dt(\om-\om_{m0}).
\end{align}
\end{widetext}
This completes the derivation of Eq.~\eqref{eq:IinelMaintext} in the main text.

\section{Low-energy theory}\label{app:low_E_theory}

{In this Appendix, we first derive the effective Dirac Hamiltonian as the low-energy theory of the lattice Kitaev model. Then, we employ the wavefunction ansatz and derive the radial functions of the Dirac Hamiltonian in the presence of the flux solenoid. The local quantum quench induced by the dynamical correlation function implies a link-variable flip, which effectively becomes a local potential in the continuum model. We derive this effective potential in the valley and sublattice space for the low-energy description.}  

\subsection{Effective Dirac Hamiltonian}
\label{app:EffectiveU1Theory}

Here, we derive the low-energy effective Dirac Hamiltonian in the continuum from the original honeycomb lattice model. To preserve the valley and sublattice space, we employ the fermion-doubling approach \cite{Zuber1977,Halasz2016} that rewrites the Hamiltonian into complex fermions with the sublattice index, and then expand it around the Dirac points $\bK$ and $\bK'$. The two-site unit cell is defined on the $J_z$ link and the Bravais lattice vectors $\ba_1 = \left(\frac{1}{2},\frac{\sqrt{3}}{2}\right)$ and $\ba_2 = \left(\frac{-1}{2},\frac{\sqrt{3}}{2}\right)$. The unit-cell translational vectors of the nearest-neighbor sites are
\begin{align}
\bdt_x = -\ba_1 \quad \bdt_y = -\ba_2,\quad \bdt_z = (0,0),
\end{align}
while for the second-neighbor sites we have
\begin{align}
\bro_1 = \ba_1-\ba_2, \quad \bro_2 = \ba_2,\quad \bro_3 = -\ba_1.
\end{align}

In the fermion-doubling approach, we introduce replica Majorana fermions $\bar{c}$ on each lattice site. The complex fermions on different sublattices are defined as
\begin{align}\label{eq:doubling_complex_fermion}
d_{j} = \frac{n_j}{2}\left(c_j+i\bar{c}_j\right), \quad n_j = \begin{cases}
1, & j \in A\\
i, & j \in B
\end{cases}.
\end{align}
Therefore, the nearest-neighbor (NN) terms of the Kitaev model in the ground-state flux sector become
\begin{align}
\begin{split}
\calH_{\rm NN} &= iJ\sum_{\langle jk\rangle}\left(c_j c_k +\bar{c}_j\bar{c}_k\right) = 2J\sum_{\langle jk\rangle}\left(d^{\dg}_{j} d^{\phantom{\dg}}_{k} + d^{\dg}_{k} d^{\phantom{\dg}}_{j}\right),
\end{split}
\end{align}
where the sublattice convention remains $j \in A$ and $k \in B$. By applying the lattice Fourier transform with $\bR_j$ as the Bravais lattice vector of site $j$ and $N_{\rm uc}$ as the unit-cell number:
\begin{align}
\fa{j} = \frac{1}{\sqrt{N_{\rm uc}}}\sum_{\bk}\fa{\bk}e^{i\bk\bR_j},
\end{align}
and separating the fermion operators on the A- and B-sublattice sites, we obtain the nearest-neighbor terms in momentum space
\begin{align}
\begin{split}
\calH_{\rm NN} = \sum_{\bk}\bg d_{\bk,A}^{\dg} & d_{\bk,B}^{\dg} \ed \bg 0 & 2S(\bk) \\ 2S^{*}(\bk) & 0 \ed \bg d_{\bk,A} \\ d_{\bk,B} \ed,
\end{split}
\end{align}
where $S(\bk) = \sum_{\bdt}J_{\bdt}e^{i\bk\cdot\bdt}$. The next-neighbor hopping is obtained similarly
\begin{align}
\calH_{\rm NNN} & =\sum_{\bk}\bg d_{\bk,A}^{\dg} & d_{\bk,B}^{\dg} \ed \bg -2\kp(\bk) & 0 \\ 0 & 2\kp(\bk) \ed \bg d_{\bk,A} \\ d_{\bk,B} \ed,
\end{align}
where $\kp(\bk) = 2\kp \sum_{\bro}\sin(\bk\cdot\mathbf{\bro})$.

The full Hamiltonian is $\calH = \calH_{\rm NN}+\calH_{\rm NNN}$, and
the energy dispersion is $\ep(\bk) = \pm 2\sqrt{|S(\bk)|^2+\kp(\bk)^2}$,  where only the positive-energy part corresponds to the Bogoliubov quasiparticle excitation of the original model.

To derive the low-energy effective model, we expand the Hamiltonian around the two Dirac points $\bK = \left( \frac{4\pi}{3},0 \right)$ and $\bK' = \left( -\frac{4\pi}{3},0\right)$,
such that $\bK\cdot\bdt_x = \bK'\cdot\bdt_y = -\frac{2\pi}{3}$ and $\bK\cdot\bdt_y = \bK'\cdot\bdt_x = \frac{2\pi}{3}$.

Around the $\bK$ and $\bK'$ points, the effective Hamiltonians are
\begin{align}
\begin{split}
&\hat{\calH}(\bK+\bp) \simeq \bg  6\sqrt{3}\kp & \sqrt{3}(-p_x+ ip_y) \\ \sqrt{3}(-p_x-ip_y) & -6\sqrt{3}\kp \ed, \\
&\hat{\calH}(\bK'+\bp) \simeq \bg  -6\sqrt{3}\kp & \sqrt{3}(p_x+ ip_y) \\ \sqrt{3}(p_x-ip_y) & 6\sqrt{3}\kp \ed.
\end{split}
\end{align}
{By defining the spinon velocity $v_s = \sqrt{3}J = \sqrt{3}$ and Haldane-like mass gap $\Dt = 6\sqrt{3}\kp$}, we obtain the low-energy effective Hamiltonian for the two valleys $\zt = \pm 1$:
\begin{align}
\begin{split}
\hat{\calH}_{\zt}(\bp)
&= -v_s(\zeta p_x\sg_x+p_y\sg_y)+\zeta \Dt \sg_z.
\end{split}
\end{align}

Therefore, the resulting continuum model is a massive Dirac Hamiltonian with two valleys.

\subsection{Solutions of the Dirac Hamiltonian}
\label{app:SolDirac}

In this Appendix, we derive the radial part of the massive Dirac Hamiltonian, Eq.~\eqref{eq:HCont}, in the presence of the flux vector potential. 
We employ the wavefunction ansatz
{
\begin{align}\label{Eq:wavefunction_ansatz}
\begin{split}
&\psi_{\mathbf{K}}(r,\phi,\zt =+1) = \bg e^{i(l-1)\phi}\,f(r) \\ ie^{il\phi}\,h(r) \ed,\\
&\psi_{\mathbf{K}'}(r,\phi,\zt =-1) = \bg e^{il\phi}\,h(r) \\ -ie^{i(l-1)\phi}\,f(r) \ed,
\end{split}
\end{align}
}
which leads to the radial differential equations:
\begin{align}
\begin{split}\label{Eq:radial_equations}
&\frac{\partial h(r)}{\partial r}+(l+\xi)\frac{h(r)}{r} = \left(\frac{E-\Dt}{v_s}\right)f(r),\\
&\frac{\partial f(r)}{\partial r}-(l+\xi-1)\frac{f(r)}{r} = -\left(\frac{E+\Dt}{v_s}\right)h(r),
\end{split}
\end{align}
{where $\xi = 0$ and $\xi = 1/2$ correspond to the zero-flux and $\pi$-flux cases, respectively.} This leads to the second-order equation for $f(r)$:
\begin{align}
f''(r)+ \frac{1}{r}f'(r)+\left[k^2-\frac{(l+\xi-1)^2}{r^2}\right]f(r) = 0,
\end{align}
where $v_s^2k^2 = E^2-\Dt^2$. This is the standard Bessel differential equation and the general solution is
\begin{align}
f(r) = A J_{\nu}(kr)+BY_{\nu}(kr), \quad \nu\equiv|l+\xi-1|.
\end{align}
For noninteger $\nu$, the second branch may equivalently be expressed in terms of $J_{-\nu}(kr)$. For $0\leq \nu < 1$, the singular branch can remain square integrable and corresponds to a different self-adjoint extension. The finite-core regularization adopted below selects the branch used in this work. Note that the singular solution diverge as $r\to 0$, and they are not square-integrable for $\nu \geq 1$. 

To solve this problem, we consider a finite range $R_s$ of the solenoid for the regularization. That means, for Region I ($r < R_s$) the particle is a free Dirac particle ($\xi = 0$), and for Region II ($r > R_s$), the particle can feel the Aharonov-Bohm flux. The spinor wavefunctions need to match at the boundary $r = R_s$, and then we take the limit $R_s\to 0$ to reach the ultra-thin flux solenoid.

First, we consider the flux-free interior region $r < R_s$.  The differential equation and the regularized solution for $f_{I}(r)$ are
\begin{align}
\begin{split}
&f_{I}'(r) - \frac{l-1}{r}f_{I}(r) = -\frac{E+\Dt}{v_s}h_I(r),\\
&f_I(r)\sim J_{\nu}(kr), \qquad \nu = |l-1|.
\end{split}
\end{align}
The negative-order Bessel functions in the general solution  $f(r)$ are discarded here due to their singular and non-integrable nature at $r = 0$. Since the order $\nu$ contains the absolute value, we separate it into two cases in the following.

For $l\geq 1$ and  $\nu = l-1$,  we quote the Bessel function property $J'_n(x) = \frac{n}{x}J_n(x)-J_{n+1}(x)$, such that
\begin{align}
\begin{split}
&f_I'(r) = k\left[\frac{l-1}{kr}J_{l-1}(kr)-J_{l}(kr)\right],\\ 
&h_I(r) = \left(\frac{v_s k}{E+\Dt}\right)J_l(kr).
\end{split}
\end{align}
At $r = R_s$, their ratio is
\begin{align}
\frac{f_{I}(R_s)}{h_{I}(R_s)} \sim \frac{J_{l-1}(kR_s)}{J_{l}(kR_s)}\sim \frac{1}{kR_s},
\end{align}
which diverges as $R_s\to 0$. This implies that $f_{I}(r)$ dominates in this branch.

For $l \leq 0$ and $\nu = 1-l$, we use $J'_n(x) = -\frac{n}{x}J_n(x)+J_{n-1}(x)$, such that
\begin{align}
h_I(r) = -\left(\frac{v_s k}{E+\Dt}\right)J_{-l}(kr) = (-1)^{l+1}\left(\frac{v_s k}{E+\Dt}\right)J_{l}(kr),
\end{align}
where we apply the property for integer-order Bessel functions: $J_{-l}(x) = (-1)^l J_l(x)$.

At $r = R_s$, their ratio is
\begin{align}
\frac{f_{I}(R_s)}{h_{I}(R_s)} \sim \frac{J_{1-l}(kR_s)}{J_{-l}(kR_s)}\sim kR_s,
\end{align}
which vanishes as $R_s\to 0$. This implies the dominance of $h_{I}(r)$ in this branch. Note that for $l = 0$, we have $f_I(r) \sim J_1(kr) \sim r$ and $h_I(r) \sim J_0(kr) \sim 1$ as $r\to 0$.

Next, we consider the exterior region $r > R_s$. The differential equations and the solutions are
\begin{align}
\begin{split}
&f_{II}'(r) - \frac{l+\xi-1}{r}f_{II}(r) = -\frac{E+\Dt}{v_s}h_{II}(r),\\
&f_{II}(r)\sim J_{\nu}(kr), \qquad \nu = |l+\xi-1|.
\end{split}
\end{align}

In the presence of a $\pi$ flux, $\nu = |l-1/2|$ and we again discuss two cases separately.

For $l\geq 1$ and $\nu = l-1/2$, we have 
\begin{align}
\begin{split}
&f_{II}'(r) = k\left[\frac{l-1/2}{kr}J_{l-1/2}(kr)-J_{l+1/2}(kr)\right],\\
& h_{II}(r) = \left(\frac{v_s k}{E+\Dt}\right)J_{l+1/2}(kr).
\end{split}
\end{align}

At $r = R_s$, their ratio is
\begin{align}
\frac{f_{II}(R_s)}{h_{II}(R_s)} \sim \frac{J_{l-1/2}(kR_s)}{J_{l+1/2}(kR_s)}\sim \frac{1}{kR_s},
\end{align}
which diverges as $R_s\to 0$. This implies the dominance of $f_{II}(r)$ in this branch.

For $l \leq 0$ and $\nu = 1/2-l$, we have
\begin{align}
\begin{split}
&f_{II}'(r) = k\left[-\frac{1/2-l}{kr}J_{1/2-l}(kr)+J_{-1/2-l}(kr)\right]\\
&h_{II}(r) = -\left(\frac{v_s k}{E+\Dt}\right)J_{-1/2-l}(kr).
\end{split}
\end{align}

At $r = R_s$, their ratio is
\begin{align}
\frac{f_{II}(R_s)}{h_{II}(R_s)} \sim \frac{J_{1/2-l}(kR_s)}{J_{-1/2-l}(kR_s)}\sim kR_s,
\end{align}
which vanishes as $R_s\to 0$. This implies the dominance of $h_{II}(r)$ in this branch. For $l = 0$, we have $f_{II}(r) \sim J_{1/2}(kr) \sim r^{1/2}$ and $h_{II}(r) \sim J_{-1/2}(kr) \sim r^{-1/2}$. It is important to note that $h_{II}(r)$ diverges as $r\to 0$, but is still square integrable since $\int |h_{II}(r)|^2 r\mathrm{d}r$ is finite. 

Comparing the ratio of A- and B-sublattice components in Region I and II, we see that they have consistent behaviors at $R_s$ for both the positive branch ($l\geq 1$) and the negative branch ($l\leq 0$). Finally, we remark that in Region II, one can in principle choose the solution $f_{II}(r) \sim J_{-\nu}(kr)$, but this would lead to an opposite spinor-component ratio and does not match the ratio of Region I at $r = R_s$. 

After taking the limit $R_s \to 0$, the entire space is covered by Region II, and the radial functions are
\begin{align}
\begin{split}
&f(r) = J_{|l+\xi-1|}(kr)\\
&h(r) = \begin{cases}\left(\frac{v_s k}{E+\Dt}\right)J_{l+\xi}(kr), & l\geq 1 \\ -\left(\frac{v_s k}{E+\Dt}\right)J_{-\xi-l}(kr), & l\leq 0   \end{cases}
\end{split}
\end{align}

To simplify the notation and incorporate the $l = 0$ component, we define the indices as functions of $l$:
\begin{align}\label{Eq:nu_eta_def}
&\nu(l,\xi) \equiv |l+\xi-1|,\quad \eta (l,\xi) \equiv \begin{cases} |l+\xi|, & l\neq 0 \\ -\xi, & l = 0\end{cases},
\end{align}
and a sign factor depending on $l$:
\begin{align}
s_l \equiv \begin{cases}+1, & l\geq 1 \\ -1, &l\leq 0\end{cases}.
\end{align}
Therefore, the radial solutions can be written concisely as
\begin{align}
\begin{split}
&f(r) = J_{\nu}(kr),\quad h(r) = s_l \left(\frac{v_s k}{E+\Dt}\right)J_{\et}(kr).
\end{split}
\end{align}

The normalized wavefunctions are obtained through the orthogonality condition
\begin{align}
\int_0^{\infty}r\mathrm{d}r\int_0^{2\pi}\mathrm{d}\phi \,\psi^{\dg}_{l}(kr)\psi_{l'}(k'r) = \frac{1}{k}\dt_{ll'}\dt(k-k').
\end{align}
We then obtain the following wavefunctions for the K and K' valleys:
\begin{widetext}

\begin{align}\label{Eq:Dirac_wavefunctions}
\begin{split}
&\psi^{(+)}_{\Phi,K} = \frac{1}{\sqrt{4\pi}}\bg \sqrt{1+\frac{\Dt}{E(k)}}\,J_{\nu}(kr)\,e^{i(l-1)\phi} \\ is_l\sqrt{1-\frac{\Dt}{E(k)}}J_{\et}(kr)e^{il\phi} \ed, \quad \psi^{(-)}_{\Phi,K} = \frac{1}{\sqrt{4\pi}}\bg \sqrt{1-\frac{\Dt}{E(k)}}\,J_{\nu}(kr)\,e^{i(l-1)\phi} \\-is_l\sqrt{1+\frac{\Dt}{E(k)}}\,J_{\et}(kr)\,e^{il\phi} \ed,\\
&\psi^{(+)}_{\Phi,K'} = \frac{1}{\sqrt{4\pi}}\bg \sqrt{1-\frac{\Dt}{E(k)}}\,J_{\et}(kr)\,e^{il\phi} \\is_l\sqrt{1+\frac{\Dt}{E(k)}}\,J_{\nu}(kr)\,e^{i(l-1)\phi} \ed, \quad \psi^{(-)}_{\Phi,K'} = \frac{1}{\sqrt{4\pi}}\bg \sqrt{1+\frac{\Dt}{E(k)}}\,J_{\et}(kr)\,e^{il\phi} \\-is_l\sqrt{1-\frac{\Dt}{E(k)}}\,J_{\nu}(kr)\,e^{i(l-1)\phi} \ed,
\end{split}
\end{align}
where the upper indices ($+$) and ($-$) denote the positive and negative energy solutions, and the lower spinor component has a sign difference between $l\geq 1$ and $l\leq0$, indicated by $s_l$. Note that when $ l = 0$ and $\xi = 1/2$, the solutions contain the negative-order Bessel function $J_{-1/2}(kr)$.

{\subsection{Effective potential due to flux-pair insertion}\label{app:effective_potential}}

{In Sec.~\ref{Sec:dynamical_spin_correlation}, we mentioned that the dynamical spin correlation function in the Kitaev model induces a flux-pair insertion that acts as a local quantum quench. Under the adiabatic approximation, the matrix element of the correlation is calculated in this sector with an additional flux pair. To incorporate this effect in the continuum model, we derive the effective flux-pair scattering potential for the Dirac Hamiltonian. Starting from the lattice model, the perturbation is a sign-reversed hopping term that creates the flux pair, and we assume that a link variable on the $z$-bond in the $\bR_0$ unit cell is flipped:}
\begin{align}
\begin{split}
\calV_{\rm tip}(\bR_0) &= -4J_z\left(\fc{\bR_0,A}\fa{\bR_0+\bdt_z, B}+\fc{\bR_0+\bdt_z,B}\fa{\bR_0,A}\right) = -\frac{4J_z}{N_{\rm uc}}\sum_{\bq,\bq'}e^{-i(\bq-\bq')\bR_0}\bg \fc{\bq,A} & \fc{\bq,B} \ed \bg 0 & e^{i\bq'\cdot \bdt_z} \\ e^{-i\bq\cdot \bdt_z} & 0  \ed \bg \fa{\bq',A} \\ \fa{\bq',B} \ed
\end{split}
\end{align}

Note that for our convention, $\bdt_z = (0,0)$ and thus the middle matrix is just $\sigma_x$ in the sublattice space. 

To see the structure in valley space, we consider the decomposition
\begin{align}
\sum_{\bq,\bq'} = \left(\sum_{\bq = \bK+\bp}\,\,\sum_{\bq'=\bK+\bp'}+\sum_{\bq = \bK'+\bp}\,\,\sum_{\bq'=\bK'+\bp'}\right)+ \left(\sum_{\bq = \bK+\bp}\,\,\sum_{\bq'=\bK'+\bp'}+\sum_{\bq = \bK'+\bp}\,\,\sum_{\bq'=\bK+\bp'}\right),
\end{align}
where the first parenthesis is for the intravalley contribution and the second parenthesis is for the inter-valley contribution. Therefore, the effective tip potential becomes
\begin{align}
\calV_{\rm tip} = -4J_z \int \mathrm{d}\br \,\, \Psi^{\dg}(\br) \bg 0 & \dt(\br-\bR_0) & 0 & e^{-i\frac{2\pi}{3}R_0}\dt(\br-\bR_0) \\ \dt(\br-\bR_0) & 0 & e^{-i\frac{2\pi}{3}R_0}\dt(\br-\bR_0) & 0 \\ 0 & e^{i\frac{2\pi}{3}R_0}\dt(\br-\bR_0) & 0 & \dt(\br-\bR_0) \\ e^{i\frac{2\pi}{3}R_0}\dt(\br-\bR_0) & 0 & \dt(\br-\bR_0) & 0 \ed \Psi(\br),
\end{align}
where $\Psi(\br) = \bg \psi_{K,A}(\br) & \psi_{K,B}(\br) & \psi_{K',A}(\br) & \psi_{K',B}(\br) \ed^T$. If we choose the impurity position such that the inter-valley phase vanishes, the resulting first-quantized potential is $\bm{V} = -4J_z (\bm{1}+\tau_x)\otimes \sg_x$, where $\bm{1}+\tau_x$ acts on the valley space and $\sg_x$ acts on the sublattice space. {This concludes the derivation of the effective potential Eq.~(\ref{eq:full_GF}) and (\ref{eq:flux_pair_potential_matrix}) of the main text.}

\end{widetext}

{\section{Analytical derivation of the retarded Green's function}\label{app:derive_GF}} 

{In this Appendix, we derive the retarded Green's function based on the wavefunction solutions of the continuum Dirac Hamiltonian. The goal is to provide explicit expressions and derivations from Eq.~(\ref{eq:GF_matrix}) to Eq.~(\ref{eq:S_and_delta}) discussed in Sec.~\ref{subsec:continnum_model_pi_flux}. We first discuss the relationship between the Majorana and complex-fermion Green's functions, and then derive the bare Green's functions in terms of Bessel and Hankel functions. Next, we focus on the on-site Green's function and derive the matrix elements in both the zero-flux and $\pi$-flux cases.}

\subsection{Relationship between Majorana and complex-fermion Green's functions}
\label{app:GFComplex}

The fermion doubling approach conveniently replaces Majorana fermion operators $c_j, \bar c_j$ by Dirac operators $d_j, d^\dagger_j$. Replica O(2) symmetry is reflected in U(1) symmetry of the complex fermions. The retarded Green's function of interest becomes
{
\begin{align}
\begin{split}
    \mathcal G^R_{jj'} (t) &\equiv - i \theta(t) \langle \{c_j(t), c_{j'}(0)\}\rangle\\
    & = n_j^{*} n_{j'} G^R_{jj'}(t) - n_jn_{j'}^{*} G^A_{j'j}(-t),
\end{split}
\end{align}
}
where $n_j$ is the sublattice factor defined in Eq.~(\ref{eq:doubling_complex_fermion}).

Here we used U(1) symmetry, resulting in the absence of Gor'kov-like anomalous Green's functions, and introduced the Green's function of complex fermions
\begin{align}
\begin{split}
    G_{jj'}^R(t) & = - i \theta(t) \langle \{d_j(t), d^\dagger_{j'}(0)\}\rangle,\\
        G_{jj'}^A(t) & = i \theta(-t) \langle \{d_j(t), d^\dagger_{j'}(0)\}\rangle.
\end{split}
\end{align}
For the on-site case, $j = j'$, we simply have $|n_j|^2 = 1$, and thus the Majorana Green's function $\mathcal G^R_{jj}(\om)$ and the Majorana local density of states $\varrho_j(\om) = -\frac{1}{\pi}\mathrm{Im}\,\mathcal{G}^R_{jj}(\om)$ in frequency space become
\begin{align}
\begin{split}
     &\mathcal G^R_{jj} (\om) = G^R_{jj} (\om) - G^A_{jj} (-\om),\\
     &\varrho_j(\om) = \rho_j(\om)+\rho_j(-\om).
\end{split}
\end{align}
Note that the full local density of states of the Majorana spectrum is particle-hole symmetric, $\varrho(-\om) = \varrho(\om)$, such that the local densities of states in the Majorana and complex-fermion representations are related by a factor of two: $\varrho(\om) = 2\rho(\om)$. In the main text, we present the results in terms of the complex-fermion LDOS because the quantity of interest is the normalized LDOS, where the numerical prefactor just cancels.  
Note that the Green's functions and LDOS in this subsection are understood to be evaluated in a fixed flux background, such that the corresponding label $\Phi$ is suppressed for clarity.

\subsection{Bare Green's functions}\label{app:bareGF}

Here we derive the bare retarded Green's function based on the wavefunctions given in Eq.~(\ref{Eq:Dirac_wavefunctions}). For example, the retarded Green's functions in the $\bK$ valley is
\begin{widetext}

\begin{align}
g_{\Phi,K}^{R}(\br,\br';\om) = \int_0^{\infty}k\mathrm{d}k\,\sum_{l=-\infty}^{\infty}\left[\frac{\psi^{(+)}_{\Phi,K,l}(\br)\left[\psi^{(+)}_{\Phi,K,l}(\br')\right]^{\dg}}{\om-E(k)+i0^{+}}+\frac{\psi^{(-)}_{\Phi,K,l}(\br)\left[\psi^{(-)}_{\Phi,K,l}(\br')\right]^{\dg}}{\om+E(k)+i0^{+}}\right],
\end{align}
and the $AA$ component is
\begin{align}
\begin{split}
G^{R,AA}_{\Phi,K}(\mathbf r,\mathbf r';\omega)
&=
\sum_{l=-\infty}^{\infty}e^{i(l-1)(\phi-\phi')}
\int_0^\infty k\mathrm{d}k\,
\frac{J_\nu(kr)J_\nu(kr')}{4\pi E(k)}
\left[
\frac{E(k)+\Delta}{\Omega-E(k)}
+
\frac{E(k)-\Delta}{\Omega+E(k)}
\right] \\
&=
\frac{\Omega+\Delta}{2\pi v_s^2}
\sum_{l=-\infty}^{\infty}e^{i(l-1)\Delta\phi}
\int_0^\infty k\mathrm{d}k\,
\frac{J_\nu(kr)J_\nu(kr')}
{q_R^2-k^2}\\
&=
-\frac{i(\Omega+\Delta)}{4v_s^2}
\sum_{l=-\infty}^{\infty}e^{i(l-1)\Delta\phi}
J_\nu(q_R r_<)H_\nu^{(1)}(q_R r_>) ,
\end{split} \label{eq:GAA}
\end{align}
where \(\Delta\phi \equiv \phi-\phi'\), $r_{<} \equiv \min(r,r')$, $r_{>} \equiv \max(r,r')$, $H_{\nu}^{(1)}$ is the Hankel function of the first kind, and $\Om = \om+i0^{+}$ is the retarded complex frequency. The radial integral is understood with the retarded branch
\(\operatorname{Im}q_R\ge0\):
\[
\int_0^\infty dk\,
\frac{kJ_\nu(kr)J_\nu(kr')}
{q_R^2-k^2}
=
-\frac{i\pi}{2}
J_\nu(q_Rr_<)H_\nu^{(1)}(q_Rr_>) .
\]

Explicitly, for the bulk-mode regime $|\om| > \Dt$, we have
\begin{align}
q_R(\omega)=
\begin{cases}
+\sqrt{\omega^2-\Delta^2}/v_s+i0^+, & \omega>\Delta\\[4pt]
-\sqrt{\omega^2-\Delta^2}/v_s+i0^+, & \omega<-\Delta
\end{cases}.
\end{align}
Throughout this work, we focus on the positive-energy bulk
continuum $\om > \Dt$, which is the frequency regime relevant to the lattice calculation.
In this regime, $q_R(\omega)=k_{\om}+i0^+$ and $k_{\om}=\sqrt{\omega^2-\Delta^2}/{v_s}>0$, and we drop the infinitesimal \(i0^+\) in the argument of the Bessel and
Hankel functions. Therefore, for this component we obtain
\begin{align}
G^{R,AA}_{\Phi,K}(\br, \br';\om) = \frac{-i(\om+\Dt)}{4v_s^2}\sum_{l} e^{i(l-1)\Dt\phi}J_{\nu}(k_{\om} r_{<})H^{(1)}_{\nu}(k_{\om} r_{>}).
\end{align}

Similarly, we can derive the other components:
\begin{align}
\begin{split}
& G^{R,BB}_{\Phi,K}(\br, \br';\om) = \frac{-i(\om-\Dt)}{4v_s^2}\sum_{l} e^{il\Dt\phi}J_{\eta}(k_{\om} r_{<})H^{(1)}_{\eta}(k_{\om} r_{>}),\\
& G^{R,AB}_{\Phi,K}(\br, \br';\om) = \frac{-ik_{\om}}{4v_s}e^{-i\phi}\sum_{l} s_l e^{il\Dt\phi}J_{\nu}(k_{\om} r_{<})H^{(1)}_{\eta}(k_{\om} r_{>}),\\
& G^{R,BA}_{\Phi,K}(\br, \br';\om) = \frac{ik_{\om}}{4v_s}e^{i\phi'}\sum_{l} s_l e^{il\Dt\phi}J_{\eta}(k_{\om} r_{<})H^{(1)}_{\nu}(k_{\om} r_{>}),\\
& G^{R,AA}_{\Phi,K'}(\br, \br';\om) = \frac{-i(\om-\Dt)}{4v_s^2}\sum_{l} e^{il\Dt\phi}J_{\eta}(k_{\om} r_{<})H^{(1)}_{\eta}(k_{\om} r_{>}),\\
& G^{R,BB}_{\Phi,K'}(\br, \br';\om) = \frac{-i(\om+\Dt)}{4v_s^2}\sum_{l} e^{i(l-1)\Dt\phi}J_{\nu}(k_{\om} r_{<})H^{(1)}_{\nu}(k_{\om} r_{>}),\\
& G^{R,AB}_{\Phi,K'}(\br, \br';\om) = \frac{-ik_{\om}}{4v_s}e^{i\phi'}\sum_{l} s_{l} e^{il\Dt\phi}J_{\eta}(k_{\om} r_{<})H^{(1)}_{\nu}(k_{\om} r_{>}),\\
& G^{R,BA}_{\Phi,K'}(\br, \br';\om) = \frac{ik_{\om}}{4v_s}e^{-i\phi}\sum_{l} s_{l} e^{il\Dt\phi}J_{\nu}(k_{\om} r_{<})H^{(1)}_{\eta}(k_{\om} r_{>}).
\end{split}
\end{align}

\end{widetext}

\subsection{Bare on-site Green's function for zero-flux and $\pi$-flux cases}\label{app:GF_components}

Here we explicitly derive the bare on-site Green's function matrices shown in Eq.~(\ref{Eq:onsite_GF_matrix}). By setting $r = r'$ and $\phi = \phi' = 0$  and defining the components of the bare Green's function matrix as
\begin{align}
\begin{split}
&g_{\Phi,K}(\br,\br;\om) = \bg a_{\Phi} & b_{\Phi} \\ c_{\Phi} & d_{\Phi} \ed, \quad g_{\Phi,K'}(\br,\br;\om) = \bg a'_{\Phi} & b'_{\Phi} \\ c'_{\Phi} & d'_{\Phi} \ed,
\end{split}
\end{align}
and in the $\ket{\pm}$ basis
\begin{align}
&g_{\Phi,\pm} = \frac{g_{\Phi,K}\pm g_{\Phi,K'}}{2} = \frac{1}{2}\bg a_{\Phi}\pm a'_{\Phi} & b_{\Phi} \pm b'_{\Phi} \\ c_{\Phi} \pm c'_{\Phi} & d_{\Phi} \pm d'_{\Phi} \ed.
\end{align}

Next, we define $x \equiv k_{\om}r$ and note that the Hankel functions can be rewritten as $H^{(1)}_{\chi}(x) = J_{\chi}(x) + iY_{\chi}(x)$. We further define the product of Bessel and Hankel functions as
\begin{align}
\begin{split}
&N_{\chi}(x) = J^2_{\chi}(x)+iJ_{\chi}(x)Y_{\chi}(x),\\
&M_{\chi,\chi'}(x) = J_{\chi}(x)J_{\chi'}(x)+iJ_{\chi}(x)Y_{\chi'}(x), 
\end{split}
\end{align}
such that the matrix elements are
\begin{align}
\begin{split}
&a_{\Phi} = d'_{\Phi} = \frac{-i(\om+\Dt)}{4v_s^2}\sum_l N_{\nu}(x),\\
&b_{\Phi} = -c'_{\Phi} = \frac{-ik_{\om}}{4v_s}\sum_l s_{l}M_{\nu,\eta}(x),\\
&c_{\Phi} = -b'_{\Phi} = \frac{ik_{\om}}{4v_s}\sum_l s_l M_{\eta,\nu}(x),\\
&d_{\Phi} = a'_{\Phi} = \frac{-i(\om-\Dt)}{4v_s^2}\sum_{l}N_{\eta}(x).
\end{split}
\end{align}
Note that the indices $\nu$ and $\et$ contain the flux variable $\xi$, as defined in Eq.~(\ref{Eq:nu_eta_def}). In the following, we discuss the structure of the matrix elements in the zero-flux and the $\pi$-flux cases separately. 

{In the zero-flux case, the indices $\nu$ and $\eta$ are both integers, such that we can define
\begin{align}
\begin{split}
S_0(x) &\equiv \sum_l N_{\nu}(x)\\
&= N_0(x)+2\sum_{n=1}^{\infty}N_n(x)\\
&= 1 + i \sum_{n= -\infty}^\infty J_n(x)Y_n(x),
\end{split}
\end{align}
and it is worth highlighting that the imaginary part is logarithmically divergent.} 
Thus, the diagonal matrix elements of $g_{0,\pm}$  become
\begin{align}
\begin{split}
&\frac{1}{2}(a_0+a_0') = \frac{1}{2}(d_0+d_0') = -\frac{i\om}{4v_s^2}S_0(x),\\
&\frac{1}{2}(a_0-a_0') = -\frac{1}{2}(d_0-d_0') = -\frac{i\Dt}{4v_s^2}S_0(x).
\end{split}
\end{align}

For the off-diagonal elements, the $l$ summation is
\begin{align}
\begin{split}
&\sum_l s_l M_{\nu,\eta}(x) = \sum_{n=0}^{\infty}\left[M_{n,n+1}(x)-M_{n+1,n}(x)\right],
\end{split}
\end{align}
and we can apply the Bessel function property $J_{n+1}(x)Y_{n}(x)-J_{n}(x)Y_{n+1}(x) = \frac{2}{\pi x}$. This leads to
\begin{align}
b_0 = c_0 = -b_0' = -c_0'  = -\frac{1}{2\pi v_s r}\left(\sum_{n=0}^{\infty}1\right),
\end{align}
and the off-diagonal matrix elements
\begin{align}
\begin{split}
&\frac{1}{2}(b_0+b_0') = \frac{1}{2}(c_0+c_0') = 0,\\
&\frac{1}{2}(b_0-b_0') =  \frac{1}{2}(c_0-c_0') = -\frac{1}{2\pi v_s r}\left(\sum_{n=0}^{\infty}1\right).
\end{split}
\end{align}

Note that the divergent sum is over partial-wave channels. For a mode with angular momentum $l$, the corresponding tangential momentum at radius $r$ is $l/r$. A microscopic momentum cutoff $\Ld \sim 1/a$ therefore implies $|l|\lesssim l_{\rm max}(r)\sim \Lambda r$. After reorganizing the \(l\)-sum into a sum over non-negative integer orders, this becomes
\begin{align}
\sum_{n=0}^{\infty}1
\rightarrow
\sum_{n=0}^{n_{\rm max}(r)}1,
\qquad
n_{\rm max}(r)\sim \Lambda r.
\end{align}

Therefore, the retarded Green's functions in the zero-flux case are:
\begin{align}
\begin{split}
&g_{0,+} = -\frac{i\om}{4v_s^2}S_0(x)\sg_0,\\
&g_{0,-} = -\frac{i\Dt}{4v_s^2}S_0(x)\sg_z -\frac{\Ld}{2\pi v_s}\sg_x. 
\end{split}
\end{align}

In the $\pi$-flux case, the indices of the Bessel functions are half-integers. We define the following terms:
\begin{align}
\begin{split}
S_{\pi}(x) &\equiv 2\sum_{n=0}^{\infty}N_{n+1/2}(x) \\
&= \frac{2}{\pi} \operatorname{Si}(2x)  + 2 i \sum_{n=0}^{\infty} J_{n + 1/2}(x)Y_{n + 1/2}(x)
\end{split}
\end{align}
and
\begin{align}
&\dt_{\pi}(x) \equiv N_{-1/2}(x)-N_{1/2}(x) = \frac{2}{\pi x}e^{i2x},
\end{align}
such that 
\begin{align}
\begin{split}
\sum_l N_{\eta}(x) &= 2\sum_{n=1}^{\infty}N_{n+1/2}(x)+N_{-1/2}(x)+N_{1/2}(x) \\ &= S_{\pi}(x)+\dt_{\pi}(x).
\end{split}
\end{align}
{Similar to $S_0(x)$, the imaginary part of $S_{\pi}(x)$ is also logarithmically divergent.
}
Therefore,
\begin{align}
\begin{split}
&a_{\pi} = d_{\pi}' = -\frac{i(\om+\Dt)}{4v_s^2}S_{\pi}(x),\\
&a_{\pi}'=d_{\pi} = -\frac{i(\om-\Dt)}{4v_s^2}\left[S_{\pi}(x)+\dt_{\pi}(x)\right],
\end{split}
\end{align}
and
\begin{align}
\begin{split}
&\frac{1}{2}(a_{\pi}+a_{\pi}')  = -\frac{i\om}{4v_s^2}\left[S_{\pi}(x)+\frac{1}{2}\left(1-\frac{\Dt}{\om}\right)\dt_{\pi}(x)\right],\\
&\frac{1}{2}(a_{\pi}-a_{\pi}') = -\frac{i\Dt}{4v_s^2}\left[S_{\pi}(x)-\frac{1}{2}\left(\frac{\om}{\Dt}-1\right)\dt_{\pi}(x)\right].
\end{split}
\end{align}

For the off-diagonal elements, we first note that
\begin{align}
\begin{split}
&b_{\pi} = -c_{\pi}' = -\frac{ik_{\om}}{4v_s}\sum_l s_l M_{\nu,\eta}(x),\\
&c_{\pi} = -b_{\pi}' = \frac{ik_{\om}}{4v_s}\sum_l s_l M_{\eta,\nu}(x). 
\end{split}
\end{align}
To evaluate the sum, we can separate out the $l = 0$ component
\begin{align}
\begin{split}
\sum_{l}s_l M_{\nu,\eta}(x) &= -M_{1/2,-1/2}(x) \\ &+ \sum_{n=0}^{\infty}\left[M_{n+1/2, n+3/2}(x)-M_{n+3/2,n+1/2}(x)\right],\\
\sum_{l}s_l M_{\eta,\nu}(x) &= -M_{-1/2,1/2}(x) \\ &+ \sum_{n=0}^{\infty}\left[M_{n+3/2, n+1/2}(x)-M_{n+1/2,n+3/2}(x)\right],
\end{split}
\end{align}
such that
\begin{widetext}
\begin{align}
\begin{split}
&\sum_{l}s_l M_{\nu,\eta}(x) + \sum_{l}s_l M_{\eta,\nu}(x) = -\left[M_{1/2,-1/2}(x)+M_{-1/2,1/2}(x)\right],\\
&\sum_{l}s_l M_{\nu,\eta}(x) - \sum_{l}s_l M_{\eta,\nu}(x) = -\left[M_{1/2,-1/2}(x)-M_{-1/2,1/2}(x)\right]+2\sum_{n=0}^{\infty}\left[M_{n+1/2,n+3/2}(x)-M_{n+3/2,n+1/2}(x)\right].
\end{split}
\end{align}
\end{widetext}

To further simplify the result, we use the Bessel function properties such that
\begin{align}
\begin{split}
&M_{1/2,-1/2}(x) + M_{-1/2,1/2}(x) = -i\frac{2}{\pi x}e^{i 2x}, \\
& \quad M_{1/2,-1/2}(x) - M_{-1/2,1/2}(x) = i\frac{2}{\pi x}.
\end{split}
\end{align}

Similarly,
\begin{align}
M_{n+1/2,n+3/2}(x) - M_{n+3/2,n+1/2}(x) = -i\frac{2}{\pi x}.
\end{align}

Therefore, we obtain the off-diagonal matrix elements
\begin{align}
\begin{split}
&\frac{1}{2}\left(b_{\pi}+b_{\pi}' \right) = -\frac{1}{2}(c_{\pi}+c_{\pi}') = \frac{1}{4\pi v_s r}e^{i2x}, \\
&\frac{1}{2}(b_{\pi}-b_{\pi}') = \frac{1}{2}(c_{\pi}-c_{\pi}') = -\frac{1}{4\pi v_s r}\left[1+2\sum_{n=0}^{\infty}1\right].
\end{split}
\end{align}
By using the same regularization as in the zero-flux case, we obtain the Green's functions in the $\pi$-flux case:
\begin{align} \label{eq:FullgpiAppendix}
\begin{split}
g_{\pi,+}  &= -\frac{i\om}{4v_s^2}\left[S_{\pi}(x)+\frac{1}{2}\left(1-\frac{\Dt}{\om}\right)\dt_{\pi}(x)\right]\sg_0  \\
&+ \frac{e^{i2x}}{4\pi v_s r} (i\sg_y),\\
g_{\pi,-} &= -\frac{i\Dt}{4v_s^2}\left[ S_{\pi}(x)-\frac{1}{2}\left(\frac{\om}{\Dt}-1\right)\dt_{\pi}(x)\right]\sg_z \\ &-\frac{1}{4\pi v_s r}\left[1 + 2\Ld r\right]\sg_x.
\end{split}
\end{align}

\section{Low-energy limit of Green's function}
\label{app:LowEnergyGF}

In this appendix, we derive the low-energy limit of the retarded Green's function near the positive-energy band edge, where the $K$-valley conduction-band states are predominantly polarized on the $A$ sublattice. The resulting Green's function for the nonrelativistic particle is the prerequisite for the semiclassical calculation discussed in the next appendix. In this limit, the energy dispersion is approximately $E(k) \simeq \Dt + \frac{v_s^2 k^2}{2\Dt}$. From Eq.~\eqref{eq:GAA}, we get
\begin{align}\label{eq:nonrel_GF}
G_{\Phi,K}^{R,AA} (\mathbf r, \mathbf r'; \omega) & \simeq \sum_{l = -\infty}^\infty e^{i (l - 1)\Dt\phi} \int \frac{k\mathrm{d}k }{2\pi} \frac{J_\nu(kr) J_\nu(kr') }{\Om - E(k)},
\end{align}
where $\Dt\phi = (\phi- \phi')$. Using the following simplified notation for the nonrelativistic particle,
\begin{align}
m = \frac{\Dt}{v_s^2}, \qquad \tilde{\om} = \om - \Dt, \qquad k_{\tilde{\om}} = \sqrt{2m\tilde{\om}},  
\end{align}
the on-site LDOS ($\br=\br'$) for $\tilde{\om} > 0$ becomes
\begin{align}
\rho^{AA}_{\Phi,K}(r,\tilde{\om}) = \begin{cases}\frac{m}{2\pi}, & \xi = 0\\
\frac{m}{\pi^2}\operatorname{Si}(2k_{\tilde{\om}}r), & \xi = 1/2
\end{cases}.
\end{align}
In the limit $k_{\tilde{\om}}r \gg 1$, we obtain the same expression for the flux-contrasted LDOS as in Eq.~(\ref{eq:deltarho}):
\begin{align}\label{eq:deltarho_nonrel}
\Dt \rho^{AA}_{K}(r,\tilde{\om}) \simeq -\frac{1}{2\pi^2} \frac{m\cos(2k_{\tilde{\om}}r)}{k_{\tilde{\om}}r}. 
\end{align}

On the other hand, the time-domain retarded Green's function can be derived from Eq.~(\ref{eq:nonrel_GF}):
\begin{align}
\begin{split}
G^{R,AA}_{\Phi,K}(\br, \br'; T) &= (-i)\Theta(T)\frac{m}{2\pi i T}\exp\left[{i\frac{m(r^2+r'^2)}{2T}}\right]\\ & \qquad \times\sum_{l}e^{i(l-1)\Dt\phi}I_{\nu}\left(\frac{-imrr'}{T}\right),
\end{split}
\end{align}
where $\Theta(T)$ is the step function and we evaluate the $k$-integral by using
\begin{align}
\begin{split}
\int_0^{\infty}k\mathrm{d}k\, & e^{-i\frac{k^2}{2m}T} \, J_{\nu}(kr)J_{\nu}(kr') = \\  &\frac{m}{iT} \exp\left[{i\frac{m(r^2+r'^2)}{2T}}\right] I_{\nu}\left(\frac{-imrr'}{T}\right).
\end{split}
\end{align}
Note that we measure the frequency relative to the positive band edge, and factor out the common phase factor $e^{-i\Dt T}$ from the time-domain Green's function.

In the zero-flux case ($\xi = 0$), the index $\nu$ for the modified Bessel function of the first kind $I_{\nu}(x)$ is just an integer, such that we can simplify the sum
\begin{align}
\sum_{l=-\infty}^{\infty}e^{i(l-1)\Dt\phi}I_{|l-1|}\left(\frac{-imrr'}{T}\right) = \exp\left(\frac{-imrr'\cos\Dt\phi}{T}\right).
\end{align}
This leads to the conventional expression of the free-particle propagator
\begin{align}
G^{R,AA}_{0,K}(\br,\br';T) = -i\Theta(T)\frac{m}{2\pi iT}\exp\left(\frac{im|\br-\br'|^2}{2T}\right).
\end{align}

For the on-site Green's function with $\br = \br'$ in both zero- and $\pi$-flux cases, we define the dimensionless variable $u \equiv mr^2/T$ (up to $\hbar = 1$) and the flux-dependent function $F_{\Phi}(u)$ such that
\begin{align}\label{eq:Parabolic}
\begin{split}
&G^{R,AA}_{\Phi,K}(\br,\br;T) = -i\Theta(T)\frac{m}{2\pi i T}F_{\Phi}\left(\frac{mr^2}{T}\right),\\
&F_{\Phi}(u) \equiv \sum_{l=-\infty}^{\infty} e^{iu}I_{\nu}(-iu) = \begin{cases} 1, & \xi = 0\\ \operatorname{erf}(e^{-i\pi/4}\sqrt{2u}), & \xi = 1/2 
\end{cases},
\end{split}
\end{align}
where $F_{0}(u)$ has no oscillatory behavior and $F_{\pi}(u)$ has the asymptotic expression for the error function when $u \gg 1$:
\begin{align}\label{eq:limiting_Fpi}
F_{\pi}(u) \simeq 1-\frac{1}{\sqrt{2\pi u}}e^{i2u} e^{i\pi/4}.
\end{align}

The flux-contrasted Green's functions in the time domain and the frequency domain are
\begin{align}
\begin{split}
&\Dt G^{R, AA}_{K}(\br,\br;T) = \Theta(T)\frac{m^{1/2}}{(2\pi)^{3/2}r}\frac{1}{\sqrt{T}}\,e^{i\left(\frac{2mr^2}{T}+\frac{\pi}{4}\right)},\\
&\Dt G^{R,AA}_{K}(\br,\br;\tilde{\om}) = e^{i\frac{\pi}{4}}\frac{m^{1/2}}{(2\pi)^{3/2}r}\int \frac{\mathrm{d}T}{\sqrt{T}}e^{i\left(\tilde{\om}T+\frac{2mr^2}{T}\right)}.
\end{split}
\end{align}
By defining the temporal phase $\Psi(T) \equiv \tilde{\om}T+\frac{2mr^2}{T}$ and applying the stationary condition $\dot{\Psi}(T_{cl}) = 0$, we obtain
\begin{align}
\tilde{\om}T_{cl} = k_{\tilde{\om}}r, \qquad \Psi(T_{cl}) = 2k_{\tilde{\om}}r. 
\end{align}
In the stationary-phase approximation, $\Psi(T) \simeq \Psi(T_{cl})+\frac{1}{2}\ddot{\Psi}(T_{cl})\tau^2$, the temporal integral gives
\begin{align}
\int_0^{\infty}\frac{\mathrm{d}T}{\sqrt{T}}e^{i\Psi(T)} \simeq \sqrt{\frac{\pi}{\tilde{\om}}}e^{i\left(2k_{\tilde{\om}}r+\frac{\pi}{4}\right)},
\end{align}
such that the flux-contrasted Green's function becomes
\begin{align}
\Dt G^{R,AA}_{K}(\br,\br; \tilde{\om}) \simeq i\frac{m}{2\pi k_{\tilde{\om}}r}e^{i2k_{\tilde{\om}}r}.
\end{align}
This result again reproduces the same flux-contrasted LDOS in Eq.~(\ref{eq:deltarho_nonrel}). 

In summary, the oscillatory phase $2mr^2/T$ is the fixed-time dynamical action of the flux-sensitive return trajectories, while $2k_{\tilde{\omega}}r$ is the corresponding fixed-energy phase selected by the temporal saddle. The topological origin of the difference between the zero- and $\pi$-flux return amplitudes is derived from the winding-sector path integral in the next appendix.

\section{Feynman path-integral approach}
\label{app:Feynman}

This appendix summarizes the Feynman path integral approach to the on-site Green's function in the presence of a central flux. We first summarize the results in App.~\ref{app:semiclassical} and subsequently discuss details: in App. \ref{app:FeynmanBerry}, the path integral in the presence of Berry curvature; in App.~\ref{app:PolarTrotter}
details about the Trotter regularization; in App.~\ref{sec:Angular}
 the angular motion and finally, App.~\ref{app:radial} the radial motion.

Throughout this appendix, we focus on the time-domain Feynman propagator $K_{\Phi}(\br_f,\br_i;T)$, which is related to the retarded Green's function in the previous appendix via $G^{R,AA}_{\Phi,K}(\br_f, \br_i; T) = -i\Theta(T)K_{\Phi}(\br_f,\br_i;T)$. As in App.~\ref{app:LowEnergyGF}, the band-edge energy is subtracted from the Hamiltonian, so that the common phase $e^{-i\Delta T}$ is factored out. Finally, to simplify the angular-momentum label, we use $\ell = l-1$ compared with the preceding sections.

\subsection{Summary of the semiclassical calculation}\label{app:semiclassical}

In this section of the appendix, we provide a summary of how Eq.~\eqref{eq:Parabolic} for the nontrivial case of a $\pi$ flux can be reached using a resummation of semiclassical trajectories encircling the solenoid, which is depicted schematically in Fig.~\ref{Fig:schematic_setup}(a). In particular, the azimuthal motion can be kept at the level of semiclassics and small fluctuations without losing the exactness of the solution. Technical details such as the regularization schemes for each of the intermediate steps are addressed in the subsequent sections.

Generally, the Feynman path integral of the transition amplitude for a particle with Berry curvature and the corresponding phase-space action have the form (see App.~\ref{app:FeynmanBerry}):
\begin{align}
&K_{\Phi}(\mathbf r_f, \mathbf r_i; T)   =    \int_{\substack{\mathbf r(-T/2) = \mathbf r _i\\ \mathbf r(T/2)= \mathbf r _f}} \mathcal D [\mathbf r , \mathbf p ] e^{i S_{ph}},\notag\\
&S_{ph}  = \int_{-\frac{T}{2}}^{\frac{T}{2}} \mathrm{d}t [ \bp\cdot \dot{\br}+\mathcal{A}(\bp)\cdot \dot{\bp} - E(\mathbf p ) - \partial_{\mathbf p }E(\mathbf p )\cdot \mathbf A (\mathbf r )].
\end{align}
\label{eq:BerryFeynman}
Here, $\mathcal A(\mathbf p )$ is the Berry connection, $\mathbf A (\mathbf r )$ the vector potential, and $E(\mathbf p)$ the dispersion relation.

However, for energies close to the bottom of the band and keeping only the long-time (low-frequency) behavior, this expression reduces to the conventional Feynman path integral for a particle with a quadratic nonrelativistic dispersion
\begin{align}
&K_{\Phi}(\mathbf r_f, \mathbf r_i; T)   =    \int_{\substack{\mathbf r(-T/2) = \mathbf r _i\\ \mathbf r(T/2)= \mathbf r _f}} \mathcal D \mathbf r e^{i S}, \notag \\
&S[\bA] = \int_{-T/2}^{T/2} \mathrm{d}t \left[\frac{m}{2} \dot{\mathbf{r}}^2 -  \mathbf{A}(\br) \cdot \dot{\mathbf{r}}\right]. 
\label{eq:PathIntegral}
\end{align}

As discussed in Sec.~\ref{Sec:continuum_model}, we use $\mathbf{A} = \frac{\Phi}{2\pi r} \hat \phi$ for a central flux as well as $\Phi \equiv 2\pi\xi$. The path integral can be solved in polar coordinates $\mathbf r(t)  = r(t) (\cos(\phi(t)), \sin(\phi(t)))$ with the action
\begin{equation}
S_{\Phi}[r,\phi] \sim \int_{-T/2}^{T/2} \mathrm{d}t \left[\frac{m}{2} \left(\dot{r}^2 + r^2 \dot \phi^2\right) - \xi \dot \phi\right]. 
\end{equation}
Note that the flux $\Phi$, defined modulo $2\pi$, plays the role of the topological theta angle. The flux-induced theta term measures the winding number $n \in \mathbb Z$ of the trajectory $\phi(t)$. The reparametrization $(x(t), y(t)) \rightarrow (r(t), \phi(t))$ requires a careful treatment of both the Jacobian associated with the coordinate change and the underlying Trotter regularization~\cite{EdwardsGulyaev1964,PeakInnomata1969,Kleinert2009}. Similarly, the symbol ``$\sim$'' indicates subtleties of the Trotter regularization; see App.~\ref{app:PolarTrotter} for these technical details.

We first integrate over $\phi$ for given, arbitrary $r(t)$ and subsequently treat the radial motion. Rather than repeating the exact evaluation of the path integral in polar coordinates~\cite{EdwardsGulyaev1964}, we here demonstrate that the summation over semiclassical trajectories yields the exact result and thereby justifies the interpretation of the return probability as a summation over braiding events, as illustrated in Fig.~\ref{Fig:schematic_setup}(a). The equation of motion is $\mathrm{d} [r^2 \dot \phi]/\mathrm{d}t = 0$ (conservation of angular momentum) and, for given winding number $n$, is solved by 
\begin{align}
    \phi(t) & = 2\pi n \frac{\int_{- T/2}^t \mathrm{d}t'/r(t')^2}{\int_{- T/2}^{T/2} \mathrm{d}t'/r(t')^2}.
\end{align}
For a trotterized version of this solution, see App.~\ref{sec:Angular}.
Inserting this solution into the action and integrating over small fluctuations around the semiclassical saddle leads to (henceforth we assume ${e^{i \phi_f} = e^{i \phi_i} = 1}
$ in this subsection)
\begin{align}
    K_{\Phi}({\mathbf r}_f, {\mathbf r}_i; T) & = \frac{1}{\sqrt{r_i r_f}} \int \mathcal Dr \left (\sum_{n \in \mathbb Z}\sqrt{\frac{\calQ}{i\pi}} e^{i\calQ n^2 - i \Phi n} \right) \notag\\
& \times e^{i\int_{- T/2}^{T/2}  \mathrm{d}t\left(\frac{m \dot r^2}{2} + \frac{1}{8 m r^2}\right)}, \label{eq:nSummationGF}
\end{align}
where $\calQ = 2\pi^2 m/ \int_{- T/2}^{T/2} \mathrm{d}t' /r(t')^2$ and we absorbed an overall prefactor into the measure. Note that this result includes careful consideration of quartic terms~\cite{EdwardsGulyaev1964} beyond the Gaussian approximation, see App.~\ref{sec:Angular}.
Poisson resummation over all topological sectors then yields
\begin{align}
    K_{\Phi}({\mathbf r}_f, {\mathbf r}_i; T) = \frac{1}{\sqrt{r_i r_f}} \int \mathcal Dr \sum_{\ell \in \mathbb Z} e^{ i\int_{- T/2}^{T/2} \mathrm{d}t  \left[\frac{m \dot r^2}{2} - \frac{1}{2mr^2}\left(\xi_{\ell}^2- \frac{1}{4}\right)\right]}
\label{eq:rPathIntegral}
\end{align}
where $\xi_{\ell} = \xi + \ell$. This result, obtained by summing over semiclassical trajectories, is precisely the same as that obtained by an exact calculation of the $\mathcal D \phi$ path-integral~\cite{EdwardsGulyaev1964}.

Finally, we briefly discuss the radial part of the path integral for $\xi = 1/2$ in App.~\ref{app:radial}. The most direct evaluation of Eq.~\eqref{eq:rPathIntegral} leads to the Bessel function appearing in Eq.~\eqref{eq:Parabolic}, as shown in Ref.~\cite{EdwardsGulyaev1964}. At the same time, we can consider the semiclassical limit for which $\ell = 0,-1$ contribute the one-dimensional free particle result $\sim \sqrt{m/(iT)}$. All other angular momentum sectors have the same classical action $S \simeq 2u$ in the limit $u = mr_i^2/T \rightarrow \infty$ with a prefactor $\sim 1/\sqrt{u}$ from the Gaussian prefactor of the reduced one-dimensional radial kernel, together with the radial measure factor $1/r_i$. Thus the semiclassical radial motion and small fluctuations reproduce the limiting behavior of Eq.~\eqref{eq:Parabolic} and Eq.~\eqref{eq:limiting_Fpi}.

\subsection{Feynman path integral for systems with Berry curvatures}
\label{app:FeynmanBerry}

In this section, we provide details about the Feynman path integral in the presence of Berry curvature. Note that in this section $\Omega$ denotes the Berry curvature and should not be confused with the complex frequency used in the preceding sections.
Also, we use $\Dt$ to denote the small time interval in Trotterization, $\Dt = T/N$, rather than the energy gap of the Dirac system, which is given by $m v_s^2$ here.

Assume a Hamiltonian $\hat H(\mathbf p )$ of a multiband system with non-degenerate eigenstates
\begin{align}
\hat H(\mathbf p ) \vert{u_{n} (\mathbf p )} \rangle = E_n(\mathbf p ) \vert{u_{n} (\mathbf p )} \rangle
\end{align}
For the Green's function in Cartesian coordinates and a starting state with orbital index $s$, which is the sublattice index in the Kitaev model, we obtain the following expression in the gauge where $\nabla \cdot \mathbf A (\mathbf r ) = 0$:
\begin{widetext}
\begin{align}
K(\mathbf r _f, s_f,\mathbf r _i, s_i; T) &= \bra{\mathbf r _f, s_f } \left (1- i\Delta :\hat H(\mathbf p +\mathbf A (\mathbf r )):  \right )^N  \ket{\mathbf r _i, s_i} \notag\\
& \simeq  \int_{\substack{\mathbf r _0 = \mathbf r _i\\ \mathbf r _N= \mathbf r _f}} \prod_{j = 1}^{N} \frac{\mathrm{d}^2 p_j \mathrm{d}^2 r_j}{(2\pi)^2} \bra{s_f}\prod_{j = 1}^{N-1}e^{i \mathbf p _{j}(\mathbf r _j - \mathbf r _{j-1})}e^{- i\Delta [\hat H(\mathbf p _{j})+\frac{\partial \hat H}{\partial \mathbf p } \vert_{\mathbf p _{j}} \cdot\mathbf A (\mathbf r _{j-1})):} \ket{ s_i} \notag\\
& \equiv \int_{\substack{\mathbf r(-T/2) = \mathbf r _i\\ \mathbf r(T/2)= \mathbf r _f}} \mathcal D [\mathbf r , \mathbf p ] e^{- i\int_{- T/2}^{T/2} \mathrm{d}t [ \dot{\mathbf p } (\mathbf r  - \mathcal A(\mathbf p )) + E(\mathbf p ) + \partial_{\mathbf p }E(\mathbf p )\cdot \mathbf A (\mathbf r )]}.
\end{align}
%\end{widetext}
This is the origin of Eq.~\eqref{eq:BerryFeynman} and we used $\mathbf 1 = \int \frac{\mathrm{d}^2 p \mathrm{d}^2r}{(2\pi)^2} e^{i \mathbf p  \cdot \mathbf r } \ket{\mathbf r } \bra{\mathbf p }$ and normal ordering $\mathbf p $ left of $\mathbf r $. We then use expansion $\mathbf 1 = \sum_n \ket{u_n(\mathbf p )} \bra{u_n(\mathbf p )}$ in band space. The dominant term comes from the band with $E_n(\mathbf p )$ close to zero (here an external potential or frequency may be kept). Note that $$\bra{u_n(\mathbf p _{j}) } \ket{ u_n(\mathbf p _{j-1})} \simeq 1 - (\mathbf p _{j} - \mathbf p _{j-1}) \bra{u_n(\mathbf p _{j}) } \ket{ \partial_{\mathbf p _j} u_n(\mathbf p _{j})},$$
is set by the Berry connection $\mathcal A(\mathbf p_j) = i\bra{u_n(\mathbf p _{j}) } \ket{ \partial_{\mathbf p _j} u_n(\mathbf p _{j})}$.
We absorbed a factor $\langle s_f\vert u_n(\mathbf p _{N-1})\rangle \langle u_n(\mathbf p _1) \vert s_i \rangle$ into the measure.

Near the bottom of a band $E(\mathbf p ) = \mathbf p ^2/2m, \partial_{\mathbf p }E(\mathbf p ) = \mathbf p /m$, we can choose a gauge $\nabla_{\mathbf p } \cdot \mathcal A(\mathbf p ) = 0$, $\Omega(\mathbf p ) \simeq \Omega = \text{ const}$. e.g. for the present case of Dirac fermions $\Omega = - 1/ (2m^2v_s^2) $. We can thus choose $\mathcal A(\mathbf p ) = * \mathbf p  \frac{\Om}{2} \equiv \hat z \times \mathbf p \frac{\Om}{2}$.
%\begin{widetext}
\begin{align}
K(\mathbf r _f, s_f,\mathbf r _i, s_i; T) &=\int_{\substack{\mathbf r (-T/2) = \mathbf r _i\\ \mathbf r (T/2)= \mathbf r _f}} \mathcal D [\mathbf r , \mathbf p ] e^{- i\int_{- T/2}^{T/2} \mathrm{d}\tau [ \dot{\mathbf p } (\mathbf r  - * \mathbf p  \frac{\Omega}{2}) +\frac{\mathbf p ^2}{2m} + \frac{\mathbf p }{m} \mathbf A (\mathbf r )]}
\end{align}
\end{widetext}
The quadratic momentum kernel to be inverted is
\begin{align}
\int \mathrm{d}\omega\frac{1}{2m}\mathbf p (- \omega) \left ( \begin{array}{cc}
1 & -  im\Omega \omega \\  im\Omega \omega & 1
\end{array}\right) \mathbf p (\omega).
\end{align}
However, the Berry-curvature term generates only higher-derivative corrections, suppressed by the ratio of the characteristic temporal frequency to the band gap. To the leading long-time order retained here, the effective action therefore reduces to Eq.~\eqref{eq:PathIntegral}.

\subsection{Polar coordinates}
\label{app:PolarTrotter}

This section contains technical details on evaluating the path integral in polar coordinates and on Trotter regularization. We consider Eq.~\eqref{eq:PathIntegral}, use $\mathbf{A} = \frac{\Phi}{2\pi r} \hat \phi$ for a central flux as well as $\xi \equiv \Phi/2\pi$.
When switching to polar coordinates $\mathbf r = r (\cos(\phi), \sin(\phi))$, two things have to be kept in mind: First, the Jacobian in the measure; second, aspects of Trotterization of the time evolution~\cite{EdwardsGulyaev1964}. This leads to
\begin{align}
K(\mathbf{r}_f, \mathbf r_i; T) = \mathcal N^{-1} \int \prod_{j=1}^{N-1} [\mathrm{d} \phi_j \mathrm{d} r_j r_j] e^{i S_{N}}
\end{align}
where $\mathcal N  = (2\pi i \Delta/m)^N$ and $N \Delta = T$. The boundary conditions on the fields are $r_0 = r_i, r_N = r_f, e^{i\phi_0} = e^{i\phi_i}, e^{i\phi_N} = e^{i\phi_f}$.

We further use 
\begin{align}
\dot{\mathbf r}^2 &\equiv \frac{(\mathbf r_j - \mathbf r_{j-1})^2}{\Delta^2}  \notag\\
&= \frac{(r_j - r_{j-1})^2 + 2 r_j r_{j -1} [1-\cos(\phi_j - \phi_{j -1})]}{\Delta^2},\\
S_{N} & = \sum_{j = 1}^N \Big [\frac{m}{2\Delta} (r_j- r_{j-1})^2\notag\\
&+ \frac{m}{\Delta} r_j r_{j-1}[1 - \cos(\phi_j - \phi_{j-1})] \notag\\
& -\xi (\phi_j - \phi_{j-1}) \Big ].
\end{align}
Note that $\sum_{j = 1}^N (\phi_j - \phi_{j-1}) = \phi_N - \phi_0 \in 2\pi \mathbb Z$.

We remark that, as usual, the Euclidean time path integral is mathematically better defined as it does not require the inclusion of convergence factors. As a matter of fact, all steps outlined here can be performed for Euclidean time as well with the corresponding Minkowski-rotated result. However, to make direct contact with the remainder of the manuscript, we here present the real-time case only.

At this point, classic works in the literature~\cite{EdwardsGulyaev1964,PeakInnomata1969} perform the $\phi$ integration exactly. We move on to the semiclassical evaluation of the $\phi$  integral, which turns out to exactly coincide with the exact integral.

\subsection{Evaluation of the angular motion using semiclassical trajectories}
\label{sec:Angular}

We evaluate the $\phi_j$ integral first, i.e., in this section we assume a fixed but arbitrary trajectory in the radial direction $\{r_j\}$.

\subsubsection{Equations of Motion}

The equations of motion are 
\begin{align}
\frac{\partial S_N}{\partial \phi_j} = \frac{m}{\Delta} [r_j r_{j-1} \sin(\phi_j - \phi_{j - 1}) - r_j r_{j+1} \sin(\phi_{j+1} - \phi_{j})]  =0
\end{align}
which in the continuum becomes $\mathrm{d}[r^2 \dot \phi]/\mathrm{d}t = 0$. The solution to the equation of motion is denoted by $\varphi_j$
\begin{align}
\varphi_j - \varphi_{j -1} = \arcsin\left ( \frac{\mathcal C_n}{r_j r_{j-1}}\right).
\end{align}
$\mathcal C_n$ is a constant determined below. We can extract the $\varphi_j$ configuration by inverting
\begin{align}
\left (\begin{array}{ccccc}
1 &-1 & 0 &\dots & 0\\
0 & 1 &-1 & 0 & \vdots \\
\vdots &\ddots & \ddots & \ddots &  \vdots \\
\vdots &\ddots & 0 & 1 &-1 \\
0 &\dots &  0& 0 & 1 
\end{array} \right ) \left (\begin{array}{c}
\varphi_N \\ \varphi_{N-1} \\\vdots \\ \varphi_1\\ \varphi_0
\end{array} \right) & = \left (\begin{array}{c}
\arcsin\left ( \frac{\mathcal C_n}{r_N r_{N-1}}\right)\\ \arcsin\left ( \frac{\mathcal C_n}{r_{N-1} r_{N-2}}\right) \\\vdots \\ \arcsin\left ( \frac{\mathcal C_n}{r_2 r_{1}}\right)\\ \arcsin\left ( \frac{\mathcal C_n}{r_1 r_{0}}\right)
\end{array} \right). \label{eq:TrafoMatrix}
\end{align}
Note that
\begin{align}
\left (\begin{array}{ccccc}
1 &-1 & 0 &\dots & 0\\
0 & 1 &-1 & 0 & \vdots \\
\vdots &\ddots & \ddots & \ddots &  \vdots \\
\vdots &\ddots & 0 & 1 &-1 \\
0 &\dots &  0& 0 & 1 
\end{array} \right ) ^{-1} = \left (\begin{array}{ccccc}
1 &1 & 1 &\dots & 1 \\
0 & 1 &1  & \dots & 1\\
\vdots &\ddots & \ddots & \ddots &  \vdots \\
\dots & 0 & 0 & 1 &1 \\
\dots & 0 & 0& 0 & 1 
\end{array} \right )
\end{align},
so that we find
\begin{equation}
{\varphi_j = \sum_{k = 1}^{j} \arcsin\left ( \frac{\mathcal C_n}{r_k r_{k-1}} \right ).}
\end{equation}
Note that neither $\phi_0$ nor $\phi_N$ is an integration variable. We fixed $\phi_0 = 0 + \mathcal O(1/N)$ without loss of generality. The constant $\mathcal C_n$ is fixed by the boundary condition $\sum_{j=1}^{N}\phi_j - \phi_{j -1} = 2\pi n$. To leading order as $N \rightarrow \infty$, we obtain
\begin{align}
\mathcal C_n \simeq	 2\pi n \left ( \sum_{j =1}^{N} \frac{1}{r_j r_{j-1}}\right )^{-1}.
\end{align}

\subsubsection{Classical action and leading corrections}

We expand about the classical solution
\begin{align}
\phi_j = \varphi_j + e^{i \pi/4}\delta_j,
\end{align}
where we used the direction of steepest descent at a 45$^\circ$ angle and keeping $\delta_j \in \mathbb R$ where $j = 1,\dots, (N-1)$ and $\delta_{0} = \delta_N = 0$ is fixed.
The action can be written as $S = S_{\rm kin} + S_{\rm top}$ where the topological term is of course $\delta_j$ independent and 
\begin{align}
S_{\rm top} = - \xi 2\pi n =  - \Phi n.
\end{align}

We further use the relation
\begin{align}
&\cos(\arcsin(x)) = \sqrt{1-x^2} \simeq 1 - x^2/2 + \dots \notag\\
&\sin(\arcsin(x)) = x
\end{align}
and $\Delta \varphi_j  = \varphi_j - \varphi_{j -1}$ to obtain the expansion
\begin{align}
\begin{split}
[1 -& \cos(\Delta \varphi_j + e^{i \pi/4}\Delta \delta_j)] \simeq 1-\cos(\Dt\varphi_j)\\
&+ e^{i\frac{\pi}{4}}\sin(\Dt\varphi_j)\Dt\dt_j + \frac{i}{2}\cos(\Dt\varphi_j)(\Dt\dt_j)^2\\
&-\frac{1}{6}e^{i\frac{3\pi}{4}}\sin(\Dt\varphi_j)(\Dt\dt_j)^3+\frac{1}{24}\cos(\Dt\varphi_j)(\Dt\dt_j)^4,
\end{split}
\end{align}
leading to the kinetic part of the action at $N\to \infty$ (keeping only $\phi$-dependent contributions):
\begin{align}
S_{\rm kin} &= \sum_{j =1}^N \frac{m}{\Delta} r_j r_{j -1}[1 - \cos(\Delta \varphi_j + e^{i \pi/4}\Delta \delta_j)] \notag\\
& \simeq \sum_{j =1}^N \frac{m}{\Delta}r_j r_{j -1} [\frac{(\Delta \varphi_j)^2}{2} +i \frac{(\Delta\delta_j)^2}{2} + \frac{(\Delta \delta_j)^4}{24}]
\end{align}%}
The quartic term must be kept (see discussion by Edwards and Gulyaev~\cite{EdwardsGulyaev1964}) but only as a prefactor to the exponential.

We find 
\begin{align}
S_{\rm kin}(\varphi)&  = \frac{m}{2\Delta} \sum_{j = 1}^N r_j r_{j-1} [\arcsin\left ( \frac{\mathcal C_n}{r_j r_{j-1}} \right )]^2 \notag\\
& \simeq \frac{m}{2\Delta} \sum_{j = 1}^N  \frac{\mathcal C_n^2}{r_j r_{j-1}} \notag\\
& \simeq \frac{(2\pi n)^2 m}{2\Delta \sum_{j =1}^N (r_j r_{j-1})^{-1}} \equiv \calQ{n^2}
\end{align}
Here we introduced the variable
\begin{align}
\calQ = \frac{2\pi^2 m}{\Delta \sum_{j =1}^N (r_j r_{j-1})^{-1}} \simeq \frac{2\pi^2 m}{\int_{- T/2}^{T/2} \frac{\mathrm{d}t}{r^2}}.
\end{align}
The fluctuation contribution is 
%\begin{widetext}
\begin{align}
\begin{split}
\int \prod_{j = 1}^{N-1} & \frac{r_j}{\pi \Delta} \mathrm{d} \delta_j \prod_{j = 1}^N\left[ (1 +i \frac{m}{\Delta} r_j r_{j-1}\frac{\Delta \delta_j^4}{24}) e^{- \frac{m}{2\Delta} r_j r_{j-1}\Delta \delta_j^2} \right ]\\
&\simeq \frac{1}{2\pi} \sqrt{\frac{\calQ}{2\pi i}} \frac{\Delta \pi}{r_N}  \prod_{j = 1}^{N} \frac{r_j}{\pi \Delta} \sqrt{\frac{\pi {2\Delta} }{{m r_j r_{j-1}}}} e^{i \frac{ \Delta }{8mr_j r_{j - 1}} } \\
& \simeq  \frac{\Delta}{2} \sqrt{\frac{\calQ}{2 \pi i r_0 r_N}} \left (\frac{2}{m \pi \Delta} \right )^{N/2} e^{i\sum_{j = 1}^N \frac{\Delta}{8m  r_j r_{j-1}}} 
\end{split}
\end{align}
We used the fact that the determinant of the transformation matrix between $\delta_j$ and $\Delta \delta_j = \delta_j - \delta_{j-1}$, Eq.~\eqref{eq:TrafoMatrix}, is unity.

In summary, the integration of $\phi$ modes at the level of classical solutions and small fluctuations leads to
\begin{widetext}
\begin{align}
\begin{split}
{K_{\Phi}({\mathbf r}_f, {\mathbf r}_i; T)}&{ = \frac{1}{\sqrt{r_i r_f}}  \int \prod_{j = 1}^{N-1} \mathrm{d}r_j  \left (\sum_{n \in \mathbb Z}\sqrt{\frac{\calQ}{i\pi}} e^{ i\calQ n^2 - i \Phi n} \right)e^{i\sum_{j = 1}^N \left [\frac{m}{2\Delta} (r_j - r_{j-1})^2 + \frac{\Delta}{8m r_j r_{j-1}}\right ]}}\\
&
\simeq \frac{1}{\sqrt{r_i r_f}} \int \mathcal Dr \left (\sum_{n \in \mathbb Z}\sqrt{\frac{\calQ}{i\pi}} e^{ i\calQ n^2 - i \Phi n} \right)e^{ i\int_{- T/2}^{T/2}  \mathrm{d}t \left(\frac{m \dot r^2}{2} + \frac{1}{8 m r^2}\right)}.
\end{split}
\end{align}
\end{widetext}
This completes the derivation of Eq.~\eqref{eq:nSummationGF} and provides a simple interpretation of the flux-contrasted propagator: the odd-winding sectors survive because they acquire the mutual-braiding phase $-1$.
\begin{align}
\begin{split}
&K_{\Phi}(\br_f,\br_i;T) \equiv \sum_{n\in \mathbb{Z}} e^{-i\Phi n}W_n (\br_f,\br_i;T)\\
&\Dt K(\br_f,\br_i;T) = \sum_{n\in\mathbb{Z}}\left[(-1)^n -1\right]W_n(\br_f,\br_i;T).
\end{split}
\end{align}

%\subsubsection{Winding-sector decomposition}

We next use the Poisson resummation formula and assume appropriate convergence factors
\begin{align}
\begin{split}
\sum_{n \in \mathbb Z}\sqrt{\frac{\calQ}{i\pi}} e^{i \calQ n^2 - i \Phi n}  & = \sum_{\ell = -\infty}^\infty \int_{-\infty}^\infty \mathrm{d}\varphi \sqrt{\frac{\calQ}{i\pi}} e^{-2 \pi i \ell \varphi} e^{i \calQ \varphi^2 - i \Phi \varphi}\\
&= \sum_{\ell = -\infty}^{\infty} e^{-i\frac{[\Phi + 2\pi \ell] ^2}{4 \calQ}}
\end{split}
\end{align}
Note that the factor $\sqrt{\calQ/(i\pi)}$ is canceled after taking the $\varphi$-integral. We thus find
\begin{align}
{K_{\Phi}({\mathbf r}_f, {\mathbf r}_i; T)  = \frac{1}{ \sqrt{r_i r_f}} \int \mathcal Dr \sum_{\ell \in \mathbb Z} e^{ i\int_{- T/2}^{T/2} \mathrm{d}t \left[ \frac{m \dot r^2}{2} - \frac{1}{2mr^2}\left(\xi_{\ell}^2- \frac{1}{4}\right)\right]}}.
\label{eq:rPathIntegralApp}
\end{align}

In summary, this completes the derivation of Eq.~\eqref{eq:rPathIntegral}. We reproduced the results of Ref.~\cite{EdwardsGulyaev1964} by summing over the semiclassical solutions and keeping the leading Gaussian fluctuations around them.

\subsection{Radial motion}
\label{app:radial}

We next consider the radial motion in the case $\xi = 1/2$, such that $\left(\xi_{\ell}^2-1/4\right)$ becomes $\ell(\ell+1)$. While the exact solution of the Schr\"odinger equation corresponds to the Bessel function, we here discuss the approximate solution.

Clearly, in the $\ell  = 0$ and $\ell = -1$ sectors, the motion corresponds to a free particle. We thus concentrate on the contribution from $\ell \neq 0, -1$. In this case, the equation of motion and its solution are
\begin{align}
m \ddot r = \frac{\ell(\ell+1)}{m r^3}, \qquad  r(t) = \sqrt{r_0^2 + (vt)^2},
\end{align}
where $ v^2 = \frac{{\ell(\ell+1)}}{m^2 r_0^2}$ is the characteristic speed scale of a radial saddle trajectory at fixed propagation time T and $r_0 \leq r(t)$ is the turnaround point (an optimization parameter, see below). Note that $r_i =  \sqrt{r_0^2 + (v T/2)^2} $.

Inserting this solution into the action yields
\begin{align}
    S &= \int_{- T/2}^{T/2} \mathrm{d}t \left [ {m {\dot r}^2} - \frac{\ell(\ell+1)}{2m r_0^2}\right] \notag\\
    &=  - T\frac{\ell(\ell+1)}{2m r_0^2} + 2\int_{r_0}^{r_i} \mathrm{d}r\left [ \sqrt{\frac{\ell(\ell+1)}{r_0^2}} \sqrt{1 - \frac{r_0^2}{r^2}}\right] \notag\\
    & = - T\frac{\ell(\ell+1)}{2m r_i^2} X^2 + 2 \sqrt{\ell(\ell+1)} \left \{\sqrt{X^2 - 1 } - \arccos\left (\frac{1}{X} \right )\right \},
\end{align}
where $X = r_i/r_0>1$. For $X \gg 1$ this can be simplified to 
\begin{align}
    S & \simeq 2 u -\pi\sqrt{\ell(\ell+1)} - \frac{\ell(\ell+1)}{2u} \left (X-\frac{2u}{\sqrt{\ell(\ell+1)}} \right)^2.
\end{align}
So the assumption of large $X$ is valid for $u \equiv m r_i^2/T \gg \sqrt{\ell (\ell+1)}$ and all reflected partial waves share the leading fixed-time dynamical phase of the classical action $2u$, yet the extremum is very shallow. 
In addition, in this limit we have
\begin{align}
X \simeq \frac{2u}{\sqrt{\ell(\ell+1)}}, \qquad v \simeq \frac{2r_i}{T},
\end{align}
and if we substitute the temporal saddle derived in App.~\ref{app:LowEnergyGF}, $T_{cl} = k_{\tilde{\om}}r/\tilde{\om}$, the corresponding speed becomes simply the group velocity of the particle $v \simeq k_{\tilde{\om}}/m$.

Next we consider the Gaussian fluctuations on top of the saddle point $t_0 = r_0/v$ (note that $t_0$ depends on $\ell$)
\begin{align}
\begin{split}
    \delta S &= \frac{m}{2} \int^{T/2}_{-T/2} \mathrm{d}t \left[ \delta \dot r^2 - \frac{3t_0^2}{(t_0^2 + t^2)^2} \delta r^2\right]\\
    &= \frac{m}{2}\int^{T/2}_{-T/2} \mathrm{d}t \, \dt r \calP_{\ell} \dt r,
\end{split}
\end{align}
where $\calP_{\ell} \equiv \calP_0 - \frac{3t_0^2}{(t_0^2+t^2)^2}$ and $\calP_0 \equiv -\partial_t^2$.
Thus, we find that the fixed-$\ell$ ($\ell \neq 0,-1$) reflected contribution to the propagator takes the form
\begin{align}
    K_{\pi,\ell}^{\rm ref}(\mathbf r_i, \mathbf r_i; T) &\sim  \frac{1}{r_i} K^{1D}_{0}(r_i, r_i; T)\left[\frac{\operatorname{Det}\calP_{\ell}}{\operatorname{Det}\calP_0}\right]^{-1/2}  e^{i 2 u}
\end{align}
where $K^{1D}_{0}(r_i, r_i; T) \sim \sqrt{m/(iT)} $ is the free propagator for one-dimensional motion. We further use the Gel'fand-Yaglom theorem to evaluate the ratio of fluctuation determinants
\begin{align}
\frac{\operatorname{Det}\calP_l}{\operatorname{Det}\calP_0} = \frac{2-X^2}{X^2},
\end{align}
which is order one in the large-$X$ limit. The total flux-sensitive semiclassical contribution is
\begin{align}
K_{\pi}^{\rm ref} = K^{\rm ref}_{\pi,0} + K^{\rm ref}_{\pi,-1} + \sum_{\ell\neq0,-1} K_{\pi,\ell}^{\rm ref}.
\end{align}
Therefore, the full coefficient and phase require the coherent partial-wave sum, including the $\ell = 0, -1$ sectors. However, for the large-$X$ and large-$u$ limits, we can estimate
\begin{align}
\Dt K(\br_i,\br_i;T) \sim \frac{1}{r_i}K^{1D}_0(r_i,r_i;T) \, e^{i2u} \sim \frac{m}{T}\frac{e^{i2u}}{\sqrt{u}},
\end{align}
such that the semiclassical radial analysis reproduces the $u^{-1/2}$ envelope and the leading fixed-time dynamical phase $2u$, consistent with the exact result in Eq.~\eqref{eq:Parabolic} and \eqref{eq:limiting_Fpi}.

\bibliography{tunneling_refs}
\end{document}